\documentclass[aps,prd,preprintnumbers,groupedaddress,showpacs,superscriptaddress,manuscript,floatfix]{revtex4-1}
\usepackage{epsfig}
\usepackage{latexsym}
\usepackage{graphicx}
\usepackage{bm}
\usepackage{longtable}
\usepackage{color}
\usepackage{ulem}
\usepackage{amssymb,amsmath}

\newcommand{\kslash}{k\kern-1ex /}
\newcommand{\pslash}{p\kern-1ex /}
\newcommand{\qslash}{q\kern-1ex /}
\newcommand{\lslash}{l\kern-1ex /}
\newcommand{\sslash}{s\kern-1ex /}
\newcommand{\Dslash}{D\kern-1.2ex /}

\newcommand{\beqa}{\begin{eqnarray}}
\newcommand{\eeqa}{\end{eqnarray}}

\newcommand{\bd}{\begin{description}}
\newcommand{\ed}{\end{description}}

\newcommand{\ben}{\begin{eqnarray}}
\newcommand{\een}{\end{eqnarray}}

\def\lsim{\raise0.3ex\hbox{$<$\kern-0.75em\raise-1.1ex\hbox{$\sim$}}}
\def\gsim{\raise0.3ex\hbox{$>$\kern-0.75em\raise-1.1ex\hbox{$\sim$}}}
\def\simgt{\rlap{\lower 3.5 pt\hbox{$\mathchar \sim$}}\raise 1.0pt \hbox {$>$}}
\def\simlt{\rlap{\lower 3.5 pt\hbox{$\mathchar \sim$}}\raise 1.0pt \hbox {$<$}}

\begin{document}

\bibliographystyle{apsrev}

\preprint{UTCCS-P-79, UTHEP-668}

\title{Study of quark mass dependence of binding energy for light nuclei in 2+1 flavor lattice QCD}

\author{Takeshi Yamazaki}
\affiliation{Faculty of Pure and Applied Sciences,
University of Tsukuba, Tsukuba, Ibaraki 305-8571, Japan}
\affiliation{Center for Computational Sciences, University of Tsukuba,
Tsukuba, Ibaraki 305-8577, Japan}
\affiliation{RIKEN Advanced Institute for Computational Science,
Kobe, Hyogo 650-0047, Japan}
\author{Ken-ichi Ishikawa}
\affiliation{Department of Physics,
Hiroshima University,
Higashi-Hiroshima, Hiroshima 739-8526, Japan}
\affiliation{RIKEN Advanced Institute for Computational Science,
Kobe, Hyogo 650-0047, Japan}
\author{Yoshinobu Kuramashi}
\affiliation{Faculty of Pure and Applied Sciences,
University of Tsukuba, Tsukuba, Ibaraki 305-8571, Japan}
\affiliation{Center for Computational Sciences, University of Tsukuba,
Tsukuba, Ibaraki 305-8577, Japan}
\affiliation{RIKEN Advanced Institute for Computational Science,
Kobe, Hyogo 650-0047, Japan}
\author{Akira Ukawa}
\affiliation{RIKEN Advanced Institute for Computational Science,
Kobe, Hyogo 650-0047, Japan}

\date{
\today
}

\begin{abstract}

We investigate the formation of light nuclei with the nuclear
mass number less than or equal to four in 2+1 flavor QCD using
a non-perturbative improved Wilson quark and 
Iwasaki gauge actions.
The quark mass is decreased from our previous work to
the one corresponding to the pion mass of 0.30 GeV.
In each multi-nucleon channel,
the energy shift of the ground state relative to the assembly of free nucleons
is calculated on two volumes, whose spatial extents are
4.3 fm and 5.8 fm.
From the volume dependence of the energy shift, 
we distinguish a bound state of multi nucleons
from an attractive scattering state.
We find that all the ground states measured in this calculation 
are bound states.
As in the previous studies at larger $m_\pi$,
our result indicates that at $m_\pi = 0.30$ GeV
the effective interaction between nucleons in the light nuclei 
is relatively stronger than 
the one in nature,
since the results for the binding energies are larger than the experimental values
and a bound state appears in the dineutron channel, which is not observed
in experiment.
Possible sources of systematic error in this calculation
are discussed.

\end{abstract}

\maketitle

\section{Introduction}
\label{sec:introduction}

The strong interaction is the origin of the formation of nuclei.
Non-perturbative lattice QCD calculation is a powerful tool
to confirm  nucleus formation from the first
principle of the strong interaction.
The nucleus formation was examined in lattice QCD 
in Ref.~\cite{Yamazaki:2009ua}, in which 
the binding energies for the $^4$He and $^3$He nuclei were
calculated in quenched QCD at a heavy quark mass corresponding to the pion mass $m_\pi=0.80$ GeV.
In this calculation, a multi-nucleon bound state was identified 
by the volume dependence of energy shift of the ground state relative to the assembly of free nucleons.
This study was followed by calculations in
$N_f = 3$ QCD at $m_\pi = 0.81$ GeV~\cite{Beane:2012vq}
and $2+1$ QCD~\cite{Yamazaki:2012hi} at $m_\pi = 0.51$ GeV.
The $^4$He nucleus formation was also reported in
a different approach using the two-nucleon potential calculation in
$N_f = 3$ QCD at $m_\pi = 0.47$, 1.02, and 1.17 GeV~\cite{Inoue:2011ai}.  
The binding energy reported is much smaller than those
in Refs.~\cite{Yamazaki:2009ua,Beane:2012vq,Yamazaki:2012hi}.
The authors in Ref.~\cite{Yamazaki:2009ua} also made the first
systematic study of the volume dependence of the energy shift for
the two-nucleon states in the spin triplet $^3$S$_1$
and singlet $^1$S$_0$ channels in quenched QCD\cite{Yamazaki:2011nd}.
This work was extended to the full QCD case in Refs.~\cite{Beane:2011iw,Beane:2012vq,Yamazaki:2012hi}; 
The volume dependence was not examined in the earlier studies
of these channels~\cite{Fukugita:1994ve,Beane:2006mx}.
The calculation in $N_f = 2 + 1$ QCD 
at $m_\pi=0.39$ GeV~\cite{Beane:2011iw}
was not conclusive of bound state formations in both channels due to
large errors of the energy shift.
Other calculations~\cite{Yamazaki:2011nd,Beane:2012vq,Yamazaki:2012hi}, on the other hand,  
concluded that there is a nucleus in each channel. 
The latter results conflict with the one from 
the two-nucleon potential calculation~\cite{Ishii:2013ira}.

The results obtained from calculation of the energy shift seem to indicate 
that the effective interaction among nucleons
seems relatively stronger,
compared to the kinetic energy of the nucleons,
than those in nature.  
Indeed, in the calculations done to date,  
the binding energies for the $^3$He and deuteron are clearly larger than
the experimental values; also
there is a bound state in the dineutron channel ($^1$S$_0$ channel), 
which has not been observed in nature.
A possible explanation of the discrepancy between the lattice QCD results and
experiment is the heavier $u,d$ quark masses employed in the calculations.
If this is the case, at the physical quark mass
the binding energies would agree with those in the nature,
and the bound state in the $^1$S$_0$ channel would disappear.
In order to check this scenario,
calculations at lighter quark masses than those 
employed in the previous calculations are necessary.
In this work, we extend our previous calculation~\cite{Yamazaki:2012hi}
at $m_\pi = 0.51$ GeV in 2+1 flavor QCD 
to a smaller quark mass of $m_\pi = 0.30$ GeV.
We investigate whether the light nuclei ($^4$He, $^3$He,
deuteron, and dineutron), which were
observed in the previous calculations~\cite{Yamazaki:2009ua,Yamazaki:2011nd,Beane:2012vq,Yamazaki:2012hi}, 
are formed or not at this quark mass.

This paper is organized as follows. In Sec.~\ref{sec:details} we explain  
details of calculation including 
the simulation parameters for gauge configuration generation
and the calculation method for the multi-nucleon channels. 
Section~\ref{sec:results} presents the results for the $^4$He, $^3$He,
deuteron ($^3$S$_1$), and dineutron ($^1$S$_0$) channels. 
Comparisons of our results with those 
in the previous studies are also discussed. 
Conclusions and discussions are given in Sec.~\ref{sec:conclusion}.

\section{Simulation details}
\label{sec:details}
\subsection{Simulation parameters}
\label{sec:sim_para}

For  gauge configuration generation in  2+1 flavor QCD,
we employ the Iwasaki gauge action~\cite{Iwasaki:2011jk} and 
a non-perturbative $O(a)$-improved Wilson quark action.
The bare coupling is fixed at $\beta = 1.90$ 
for which we use $c_{\rm SW} = 1.715$~\cite{Aoki:2005et}.
The lattice spacing is $a=0.08995(40)$ fm, 
corresponding to $a^{-1} = 2.194(10)$ GeV, which was
determined by $m_\Omega=1.6725$ GeV~\cite{Aoki:2009ix}.
We choose two lattice sizes, 
$L^3\times T = 48^3 \times 48$ and $64^3 \times 64$, 
to investigate the spatial volume dependence of the energy 
shift between the multi-nucleon ground state and the free nucleons.
The physical spatial extents are 4.3 and 5.8 fm, respectively.
We choose the hopping parameters
$(\kappa_{ud},\kappa_s) = (0.1376314,0.1367299)$
to obtain $m_\pi = 0.30$ GeV and 
the physical strange quark mass, which are
determined by an analysis with the results for $m_\pi$ and $m_s$ 
obtained with the same actions but at
different hopping parameters~\cite{Aoki:2008sm,Aoki:2009ix}.

We utilize the domain-decomposed Hybrid-Monte-Carlo (DDHMC)
algorithm~\cite{Luscher:2003vf,Luscher:2005rx} with
mass preconditioning~\cite{Hasenbusch:2001ne}, 
{\it i.e.,} mass-preconditioned 
DDHMC (MPDDHMC), for the degenerate light quarks and 
the UV-filtered polynomial HMC (UVPHMC)
algorithm~\cite{Ishikawa:2006pb} for the strange quark.
In both algorithms, we use the 
Omelyan-Mryglod-Folk integrator~\cite{{Omelyan:2003om},{Takaishi:2005tz}}
in the molecular dynamics evolution.
The algorithmic details are given in Ref.~\cite{Aoki:2008sm}.
We summarize the simulation parameters in Table~\ref{tab:conf_gene} including
the block sizes and the preconditioning factor in MPDDHMC 
and the polynomial order in UVPHMC.
We take $\tau = 1$ for the trajectory length of the molecular dynamics 
in all the runs. The step sizes are
chosen such that we obtain the reasonable acceptance rates presented 
in Table~\ref{tab:conf_gene}.
In the spatial extents of 4.3 fm and 5.8 fm,
1000 and 800 trajectories are generated in four and two streams
after thermalization,
and the total lengths of trajectory for the measurement are
4000 and 1600, respectively.

We calculate correlation functions in the multi-nucleon channels
in every 10 trajectories for both volumes using
the same quark action as for the configuration generation.
The errors are estimated by jackknife analysis choosing 200 and 160
trajectories for the bin size for the smaller and larger
volumes, respectively.
Statistics is increased by repeating the measurement of 
the correlation functions with different source positions 
on each configuration.
We calculate the correlation functions not only in 
the temporal direction but also in the spatial ones with 
the use of the space-time rotational symmetry.
It allows to increase the statistics by a factor four effectively.
The parameters of the measurement, {\it e.g.},
the number of configurations and the bin sizes,
are listed in Table~\ref{tab:conf_meas}.

\subsection{Calculation method}
\label{sec:method}

We extract the ground state energy in the multi-nucleon channels 
and the nucleon mass from the correlation function,
\begin{equation}
G_{\mathcal{O}}(t) = \langle 0 | \mathcal{O}(t)
\overline{\mathcal{O}}(0) | 0 \rangle,
\label{eq:corr}
\end{equation}
with $\mathcal{O}$ being proper operators
for the $^4$He, $^3$He, $^3$S$_1$ and $^1$S$_0$ channels 
and also the nucleon $N$, which are given in the next subsection.

We define the energy shift between the multi-nucleon ground state and
free nucleons on finite volume as,
\begin{equation}
\Delta E_L=E_{\mathcal{O}}-N_N m_N,
\label{eq:delE_L}
\end{equation}
where $E_{\mathcal{O}}$ is the lowest energy level for a multi-nucleon
channel, $N_N$ is the number of nucleons in the channel,
and $m_N$ is the nucleon mass. 
This quantity is directly extracted from the ratio of the multi-nucleon 
correlation function to
the $N_N$th power of the nucleon correlation function
\begin{equation}
R(t) = \frac{G_{\mathcal{O}}(t)}{\left(G_N(t)\right)^{N_N}},
\label{eq:R}
\end{equation}
in the large time region where both correlation functions
are dominated by the ground state.
We also define an effective energy shift as
\begin{equation}
\Delta E_L^{\mathrm{eff}} = \ln \left(\frac{R(t)}{R(t+1)}\right),
\label{eq:eff_delE_L}
\end{equation}
which is utilized to investigate plateau region in later section.
Note that the definition of $\Delta E_L$ and $\Delta E_L^{\mathrm{eff}}$ 
follows that in Refs.~\cite{Yamazaki:2011nd,Yamazaki:2012hi},
while the sign convention is opposite to that in Ref.~\cite{Yamazaki:2009ua}.

There are two computational difficulties in the calculation of
$G_{\mathcal{O}}(t)$ for multi-nucleon channels.
One is a factorially large number of 
Wick contractions for quark-antiquark fields.
To overcome the difficulty, we use the reduction technique
of calculation cost proposed in our exploratory work~\cite{Yamazaki:2009ua}.
It is noted that other reduction techniques for the large number of 
Wick contractions have been proposed for the multi-meson~\cite{Detmold:2010au}
and multi-baryon~\cite{Doi:2012xd,Detmold:2012eu,Gunther:2013xj} channels.
Another problem is an exponential increase of statistical errors with
atomic mass number.
For this difficulty, we carry out measurements as much as possible
using multiple source points.
The number of measurements are a factor twelve and five times larger than
those in the previous calculation of $m_\pi = 0.51$ GeV~\cite{Yamazaki:2012hi}
for 4.3 fm and 5.8 fm spatial extents, respectively.

Another difficulty in the nucleus calculation is 
to distinguish a multi-nucleon bound state from
an attractive scattering state in a finite volume~\cite{{Luscher:1986pf},{Beane:2003da},{Sasaki:2006jn}}.
This problem is handled by studying the volume dependence
of the measured $\Delta E_L$ as in Refs.~\cite{Yamazaki:2009ua,Yamazaki:2011nd}.
While the energy shift of an attractive scattering state 
vanishes in the infinite volume limit as $1/L^3$~\cite{Luscher:1986pf,Beane:2007qr}, 
the physical binding energy of a bound state 
remains at a finite value in the limit.

\subsection{Interpolating operators}
\label{sec:operator}

The $u,d$ quark propagators are solved with
the periodic boundary condition 
in all of spatial and temporal directions
using an exponentially smeared source,
\begin{equation}
q(\vec{x},t) = \sum_{\vec{y}} A\, e^{-B|{\vec x} - \vec{y}|} q_0(\vec{y},t),
\label{eq:smear}
\end{equation}
for $|\vec{x}|\ne 0$, and $q(\vec{x},t) = 1$ for $|\vec{x}| = 0$,
after the Coulomb gauge fixing, where $q_0$ is the local quark field.
We choose the smearing parameters $A=0.8$ and $B=0.16$ on the two volumes
to obtain reasonable plateaus of the effective energy for 
the nucleon and ground states in the multi-nucleon channels.
The stopping condition of the quark propagator 
$\epsilon = |Dx-b|/|b| < 10^{-6}$   
is applied in both volumes to reduce the calculation time.
We have checked in a subset of the configurations that the results 
with this looser stopping condition agree with
the ones using a more stringent stopping condition $\epsilon < 10^{-14}$
in more than six digits,
even for $G_{^4{\rm He}}(t)$ at $t=12$.
The systematic error coming from the discrepancy is much smaller 
than the statistical error in the current calculation.

The interpolating operator for the proton is defined as 
\begin{equation}
p_\alpha = \varepsilon_{abc}([u_a]^tC\gamma_5 d_b)u_c^\alpha,
\label{eq:def:proton}
\end{equation}
where $C = \gamma_4 \gamma_2$ and $\alpha$ and $a,b,c$ are the Dirac and
color indices, respectively.
The neutron operator $n_\alpha$ is obtained 
by replacing $u_c^\alpha$ by $d_c^\alpha$ in Eq.(\ref{eq:def:proton}).

The $^4$He nucleus has zero total angular momentum and positive parity
$J^P = 0^+$ with the isospin $I = 0$. 
We employ the simplest $^4$He interpolating operator with  
zero orbital angular momentum $L=0$, and hence $J=S$ with $S$ being
the total spin~\cite{Beam:1967zz},
\begin{equation}
^4\mathrm{He} = 
\frac{1}{\sqrt{2}}\left( \overline{\chi}\eta - 
\chi \overline{\eta} \right),
\end{equation}
where 
\begin{eqnarray}
\chi &=&  \frac{1}{2}( [+-+-] + [-+-+] - [+--+] - [-++-] ),\\
\overline{\chi}  &=& \frac{1}{\sqrt{12}}( 
[+-+-] + [-+-+] + [+--+] + [-++-] - 2 [++--] - 2 [--++] 
),
\end{eqnarray}
with $+/-$ being up/down spin of each nucleon, and  
$\eta, \overline{\eta}$ are obtained
by replacing $+/-$ in $\chi, \overline{\chi}$ by $p/n$ for the isospin.

The $^3$He nucleus has 
$J^P=\frac{1}{2}^+$, $I = \frac{1}{2}$ and $I_z = \frac{1}{2}$.
We employ the interpolating operator in Ref.~\cite{Bolsterli:1964zz},
\begin{equation}
^3{\rm He} = 
\frac{1}{\sqrt{6}}\left(
\left|p_- n_+ p_+ \right\rangle
-
\left|p_+ n_+ p_- \right\rangle
+
\left|n_+ p_+ p_- \right\rangle
-
\left|n_+ p_- p_+ \right\rangle
+
\left|p_+ p_- n_+ \right\rangle
-
\left|p_- p_+ n_+ \right\rangle
\right).
\end{equation}

The two-nucleon operators for the $^3$S$_1$ and $^1$S$_0$ 
channels are given by 
\begin{eqnarray}
NN_{^3{\mathrm S}_1}(t) &=& \frac{1}{\sqrt{2}}\left[
p_+(t) n_+(t) - n_+(t) p_+(t)
\right],
\label{eq:def:3S1}\\
NN_{^1{\mathrm S}_0}(t) &=& 
\frac{1}{\sqrt{2}}\left[
p_+(t) p_-(t) - p_-(t) p_+(t)
\right],
\label{eq:def:1S0}
\end{eqnarray}
respectively.
In the $^3$S$_1$ channel
the operators for the other two spin components are constructed in a similar way.
We increase statistics by averaging over the three correlation
function with each spin component operator.

Using the interpolating operators above, we calculate
correlation functions in each channel.
For the source operators in all correlation functions, 
we insert the smeared quark fields of 
Eq.~(\ref{eq:smear}) for each nucleon operator located at 
the same spatial point $\vec{x}$.
Each nucleon in the sink operator, on the other hand, 
is composed of the point quark fields 
corresponding to $q_0$ in Eq.~(\ref{eq:smear}),  
and projected to zero spatial momentum. 
To save the computational cost
we use the non-relativistic quark operators, in which the Dirac index
in Eq.~(\ref{eq:def:proton}) is restricted to the upper two components
in the Dirac representation.

\section{Results}
\label{sec:results}

\subsection{Nucleon and pion masses}

\begin{figure}[!t]
\includegraphics*[angle=0,width=0.6\textwidth]{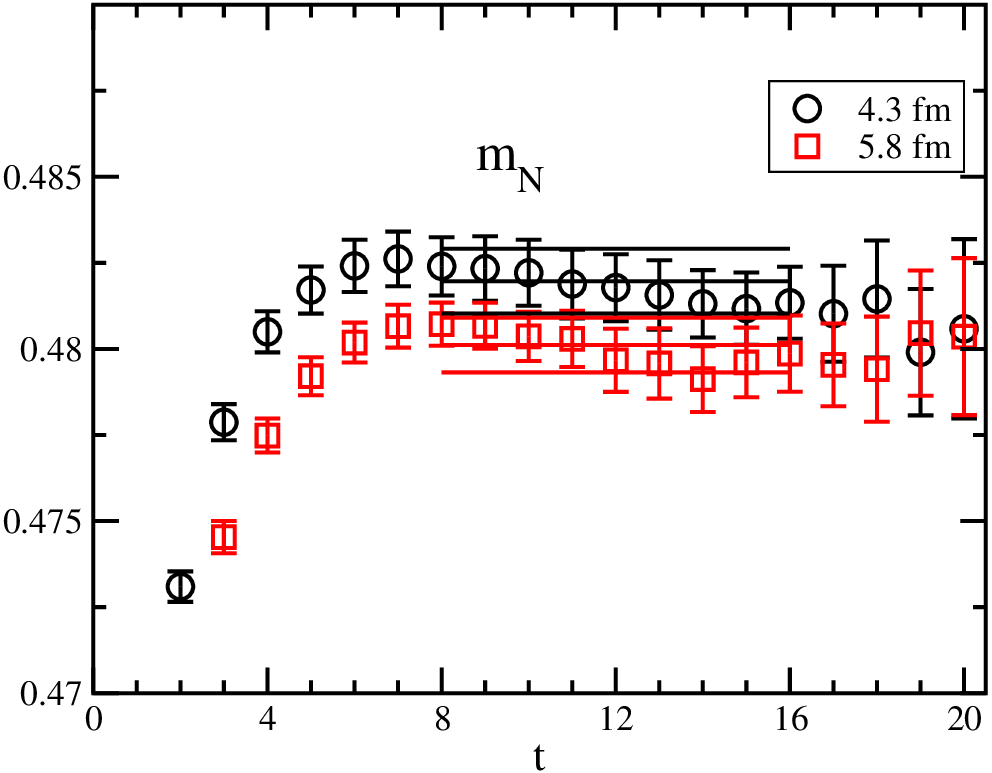}
\caption{
Nucleon effective masses on (4.3 fm)$^3$ and
(5.8 fm)$^3$ volumes in lattice unites.
Fit result with one standard deviation error band is expressed
by solid lines.
\label{fig:eff_N}
}
\end{figure}

\begin{figure}[!t]
\includegraphics*[angle=0,width=0.6\textwidth]{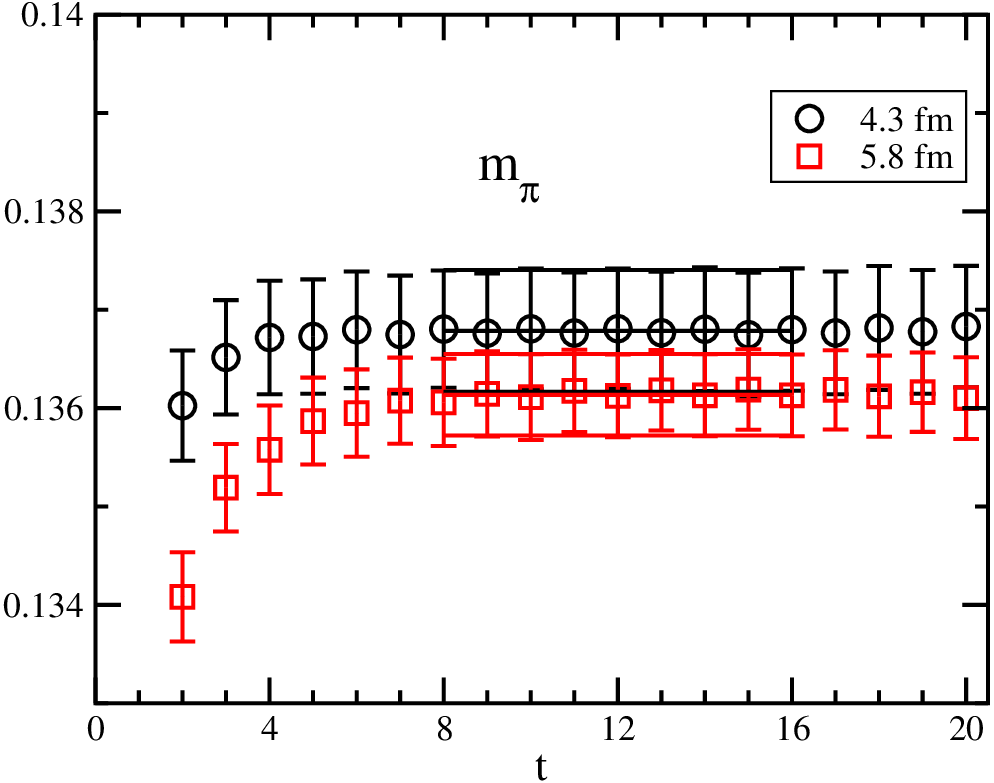}
\caption{
Same figure as Fig.~\ref{fig:eff_N}, but for
pion effective masses.
\label{fig:eff_pi}
}
\end{figure}

The results for effective $m_N$ in the two volumes are shown in 
Fig.~\ref{fig:eff_N}
together with the exponential fit result of $C_N(t)$ 
and the one standard deviation error band.
Plateaus are clearly seen for $t\ge 8$ for both volumes.
The difference of the fit results between the two volumes is 1.4 standard deviations, and hence 
statistically not very significant.  We also do not expect much finite size effect for these large volumes satisfying $m_\pi L > 6$.  
In the following sections, we therefore consider that the difference is caused by statistics,
and will not estimate the systematic error from it.
The pion effective masses in each volume
show better consistency than the nucleon mass, 
as presented in Fig.~\ref{fig:eff_pi}.
Those fit results are tabulated in Table~\ref{tab:conf_meas}.

\subsection{$^4$He channel}
\label{sec:4N}

\begin{figure}[!t]
\includegraphics*[angle=0,width=0.6\textwidth]{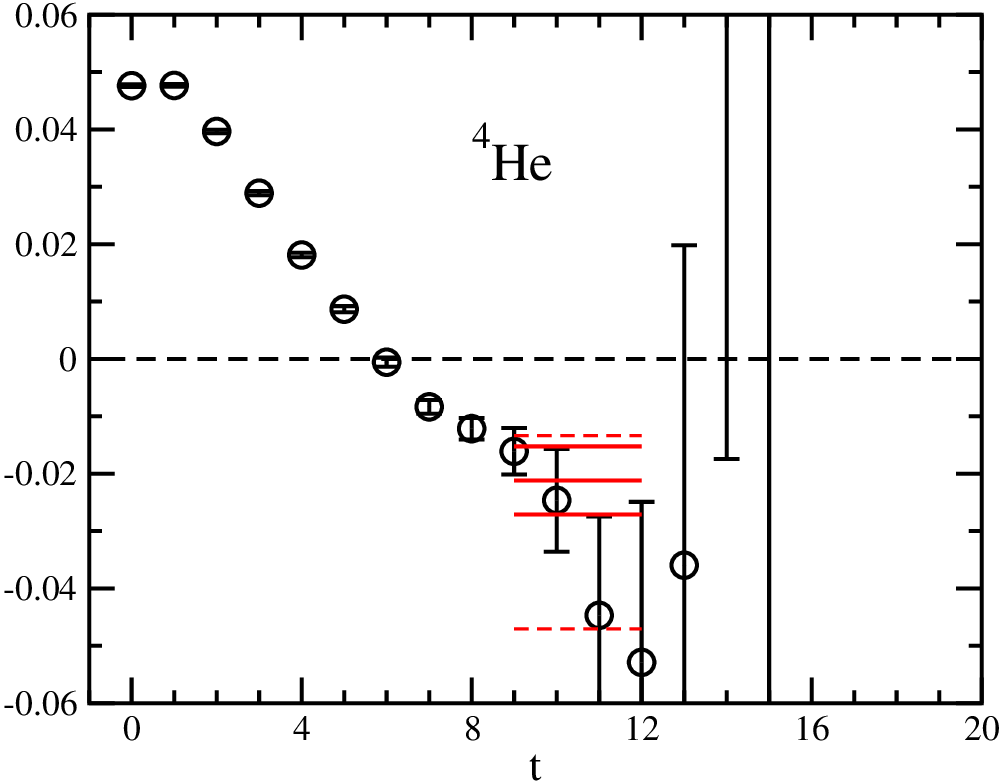}
\includegraphics*[angle=0,width=0.6\textwidth]{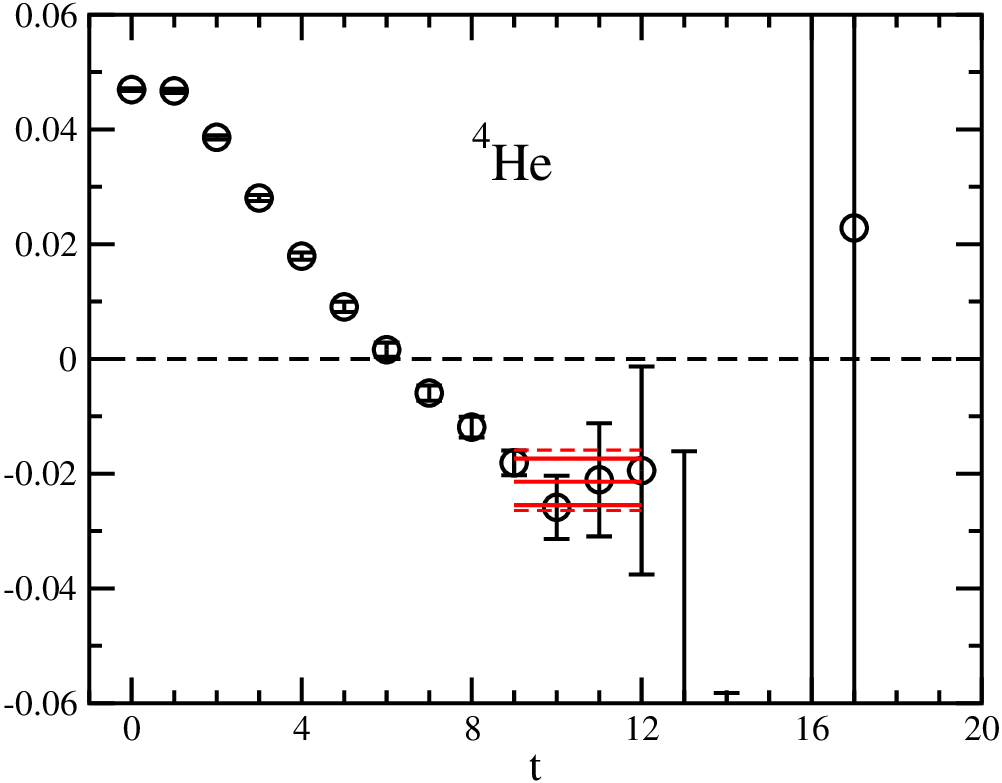}
\caption{
Effective energy shift 
$\Delta E_L^{\mathrm{eff}}$
for $^4$He channel on (4.3 fm)$^3$~(top) and (5.8 fm)$^3$~(bottom) volumes 
in lattice units.
Fit result with one standard deviation error band and 
total error including the systematic one is expressed
by solid and dashed lines, respectively.
\label{fig:eff_R_h4}
}
\end{figure}

\begin{figure}[!t]
\includegraphics*[angle=0,width=0.6\textwidth]{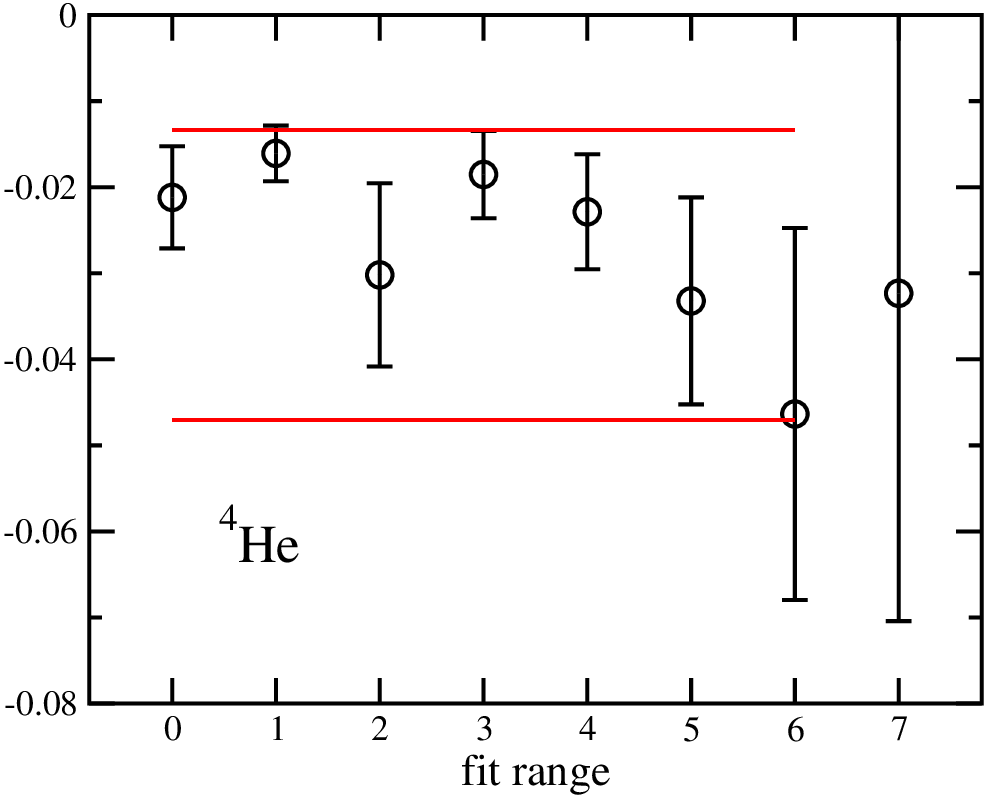}
\includegraphics*[angle=0,width=0.6\textwidth]{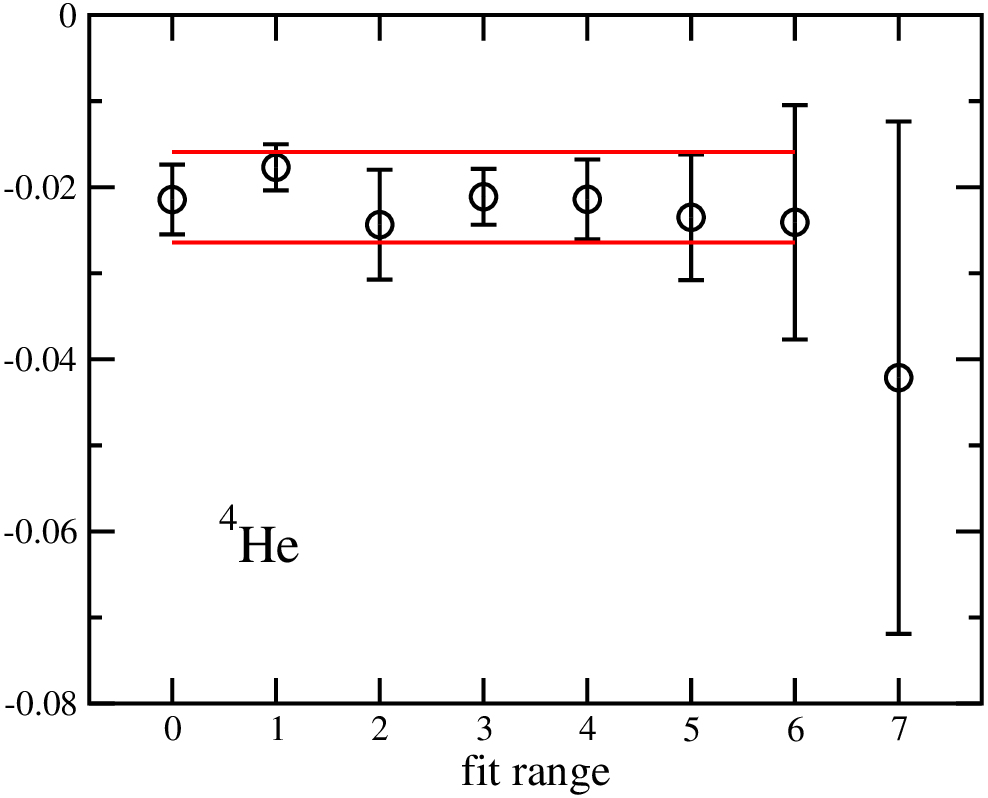}
\caption{
Fit range dependence of energy shift $\Delta E_L$ for $^4$He channel 
on (4.3 fm)$^3$~(top) and (5.8 fm)$^3$~(bottom) volumes 
in lattice units.
The horizontal axis corresponds to the fit range 
$(t_{\rm min},t_{\rm max}), 
(t_{\rm min}-1,t_{\rm max}),
(t_{\rm min}+1,t_{\rm max}),
(t_{\rm min},t_{\rm max}-1),
(t_{\rm min},t_{\rm max}+1),
(t_{\rm min}+1,t_{\rm max}+1),
(t_{\rm min}+2,t_{\rm max}+2),$ and
$(t_{\rm min}+3,t_{\rm max}+3)$ from left to right.
$t_{\rm min}$ and $t_{\rm max}$ are minimum and maximum time slices
of the fit range, respectively, whose values are explained in the text.
The total error band including the statistical and systematic is expressed
by solid lines.
\label{fig:fit_dep_h4}
}
\end{figure}

The effective energy energy shift $\Delta E^{\mathrm{eff}}_L$ defined in
Eq.~(\ref{eq:eff_delE_L}) is shown in Fig.~\ref{fig:eff_R_h4} 
for the two volumes.
Clear signals are seen for $t \le 10$,
but for larger $t$ the statistical error increases rapidly.
A plateau appears at $t=9$--12 on the larger volume,
while it is not clearly seen on the smaller volume.
More statistics is desirable for establishing a plateau in this case.
We calculate the energy shift $\Delta E_L$ in Eq.~(\ref{eq:delE_L}) by
a single exponential fit of $R(t)$ in Eq.~(\ref{eq:R})
using the same range $t = 9$--12 for the two volumes.
The systematic error is estimated from the variation of the fit results with
six different fit ranges,
where the minimum or maximum time slice is changed by $\pm 1$,
and the minimum and maximum time slices are changed by $+1$ and $+2$.
The central fit result is shown in Fig.~\ref{fig:eff_R_h4}
by solid lines with the band representing the statistical error.
The dashed lines denote the total error adding 
the statistical and systematic errors by quadrature.
In Fig.~\ref{fig:fit_dep_h4} we illustrate how we estimate 
the systematic error.  
Shown in the figure are the results of 8 fits obtained by shifting 
the fitting range as explained in the figure caption.
The horizontal band with solid lines shows the total error obtained by 
adding the statistical error and the systematic error from the 7 fits on 
the left by quadrature.
We observe that the 7 fits on the left reasonably covers the variation, 
with the 8th fit at the rightmost with the fitting range shifted by +3 from 
($t_{\rm min}, t_{\rm max}$) falling within the band of solid lines within 
one sigma.  
We therefore consider that our estimate of systematic errors is 
reasonable under the current statistics. 
The values of $\Delta E_L$ with the statistical 
and systematic errors are summarized in Table~\ref{tab:dE_He}.

\begin{figure}[!t]
\includegraphics*[angle=0,width=0.6\textwidth]{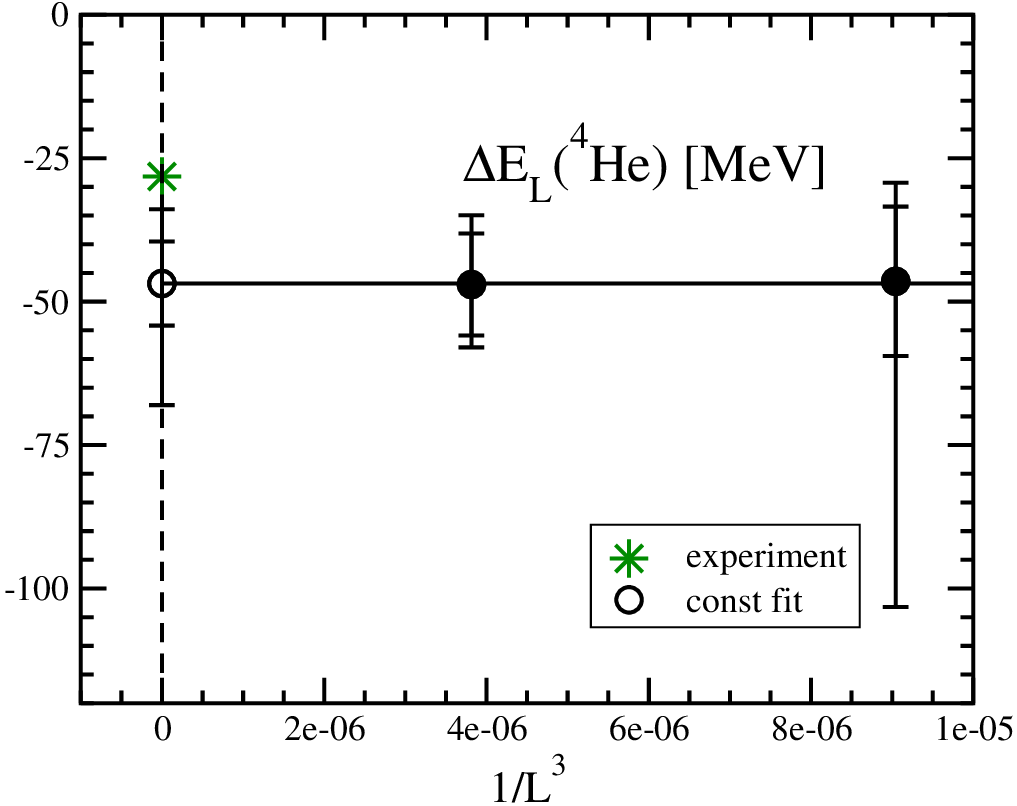}
\caption{
Spatial volume dependence of $\Delta E_L$ for $^4$He channel in MeV units.
Outer bar denotes the total error of statistical and 
systematic ones added in quadrature. Inner bar is for
the statistical error.
Constant fit result is shown by open circle symbol.
Experimental value (star) is also presented.
\label{fig:dE_h4}
}
\end{figure}

Figure~\ref{fig:dE_h4} shows the volume dependence of $\Delta E_L$
as a function of $1/L^3$.
The inner bar of each data denotes the statistical error and 
the outer bar represents the 
total error with the statistical and 
systematic ones added in quadrature.
Since the volume dependence is not large, 
we estimate the energy shift in the infinite volume limit
$\Delta E_\infty$ by a constant fit as presented by solid line
and open circle in Fig.~\ref{fig:dE_h4}.
An exponential type extrapolation, 
$\Delta E_L = \Delta E_\infty + C \exp(-C_e L)$,
cannot be carried out in this work 
due to smaller number of data than its free parameters.
The systematic error is estimated from the variation of
the central values obtained by 49 fits.
The 49 fits are constant fits with various combination of 
7$^2$ data set, where in each volume we have 7 data 
with different fit range of $R(t)$ as explained in the above.
The result of $\Delta E_\infty$ with the statistical
and systematic errors are tabulated in Table~\ref{tab:dE_He}.
From the result that $\Delta E_\infty$ is non-zero and negative,
we conclude that the ground state is bound in this channel.
The binding energy equals  $-\Delta E_\infty=47(7)(^{+20}_{-11})$ MeV
where the first and second errors are statistical and systematic, respectively.

\begin{figure}[!t]
\includegraphics*[angle=0,width=0.6\textwidth]{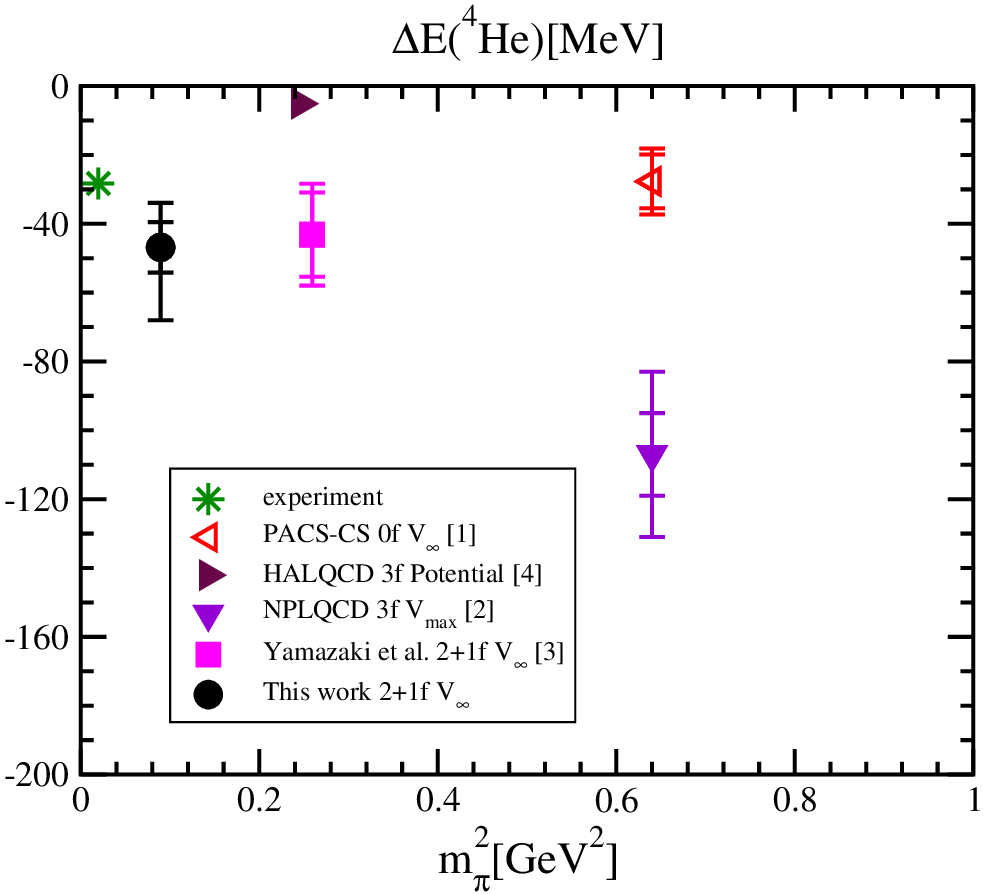}
\caption{
$m_\pi^2$ dependence of energy shift for $^4$He channel
in MeV units.
Open and closed symbols denote the quenched~\cite{Yamazaki:2009ua} 
and full QCD~\cite{Inoue:2011ai,Beane:2012vq,Yamazaki:2012hi}
results, respectively. 
The results of Refs.~\cite{Yamazaki:2009ua,Yamazaki:2012hi} 
and this work are the ones in the infinite volume limit.
The error of the result obtained from the two-nucleon potential 
was not estimated in Ref.~\cite{Inoue:2011ai}.
Experimental result (star) is also presented for comparison.
\label{fig:mdep_4He}
}
\end{figure}

The result for $-\Delta E_\infty$ is compared with the experimental
value of 28.3 MeV and with  
the previous three results~\cite{Yamazaki:2009ua,Beane:2012vq,Yamazaki:2012hi}
in Fig.~\ref{fig:mdep_4He}.
The binding energy for $m_\pi=0.3$~GeV obtained in this work is similar in magnitude 
with our previous results for
$N_f = 2+1$ at $m_\pi = 0.51$ GeV~\cite{Yamazaki:2012hi} and 
$N_f = 0$ at $m_\pi = 0.80$ GeV~\cite{Yamazaki:2011nd}.
Compared to experiment, if one used the upper total error, our current value is 
consistent within 1.5 $\sigma$.  
The result of the $N_f = 3$ calculation at $m_\pi=0.81$ GeV~\cite{Beane:2012vq} is
about three times larger than the other results.
This difference might be due to different quark masses of the calculation 
or dynamical quark effects.
On the other hand, the result obtained with the  two-nucleon potential
extracted from $N_f = 3$ calculations at $m_\pi = 0.47$ GeV~\cite{Inoue:2011ai} 
has a very small binding energy, $\Delta E = 5.1$ MeV,
compared to the other results.

\subsection{$^3$He channel}
\label{sec:3N}

\begin{figure}[!t]
\includegraphics*[angle=0,width=0.6\textwidth]{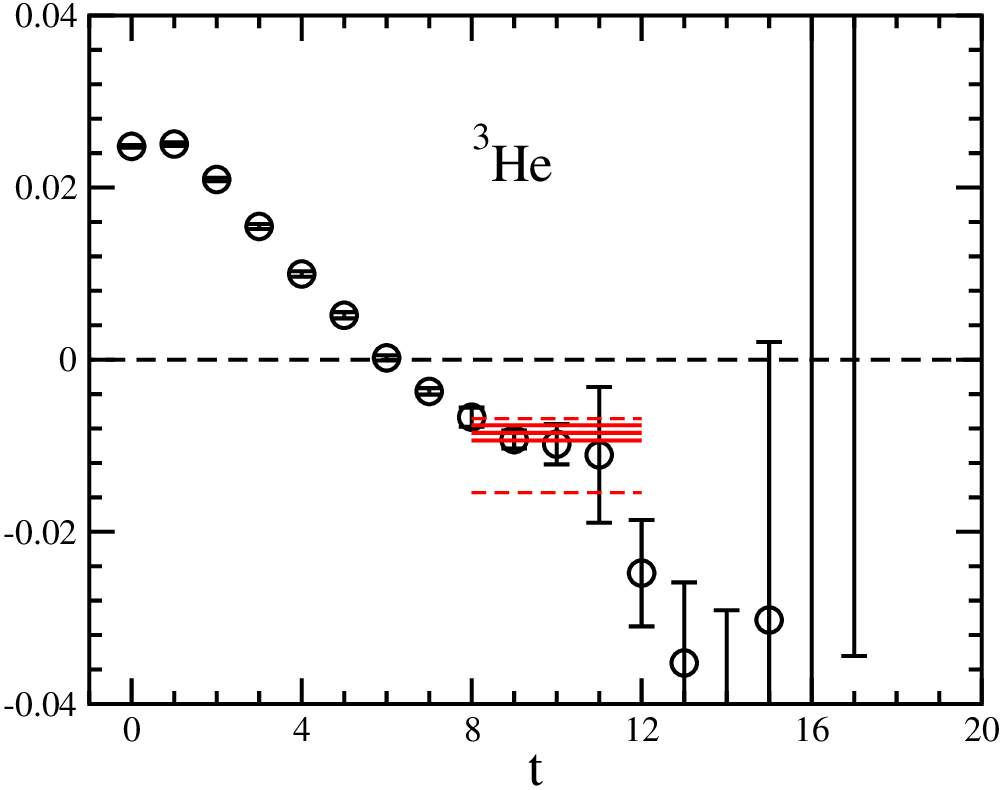}
\includegraphics*[angle=0,width=0.6\textwidth]{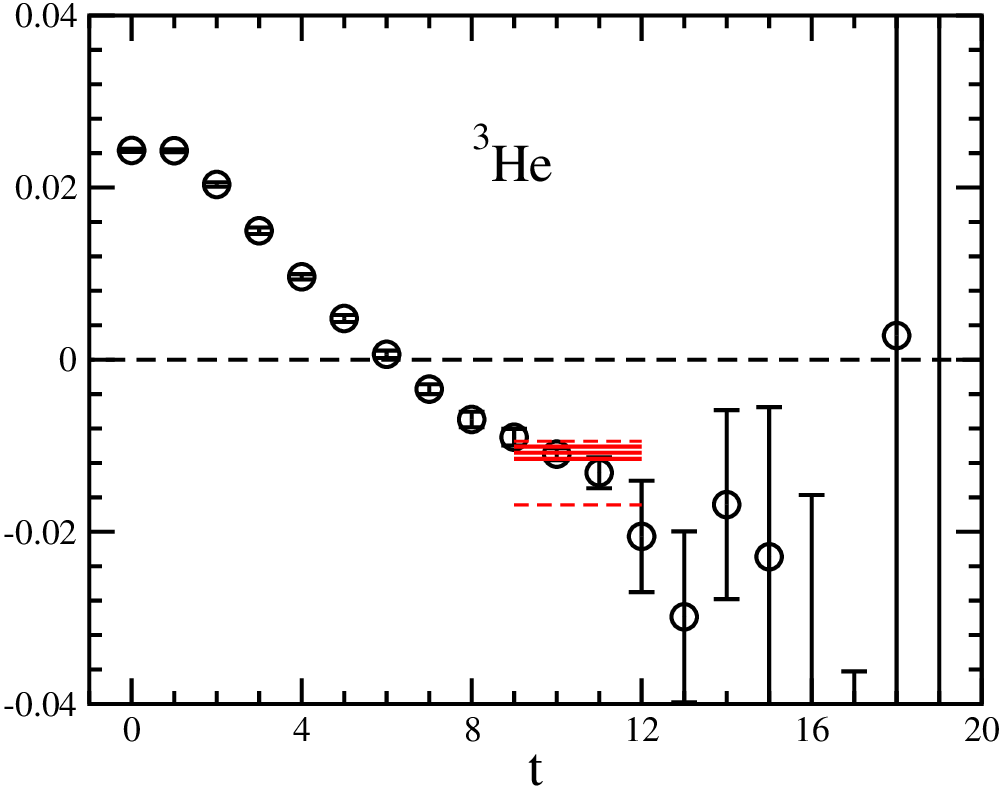}
\caption{
Same figure as Fig.~\ref{fig:eff_R_h4}, but for $^3$He channel.
\label{fig:eff_R_h3}
}
\end{figure}

\begin{figure}[!t]
\includegraphics*[angle=0,width=0.6\textwidth]{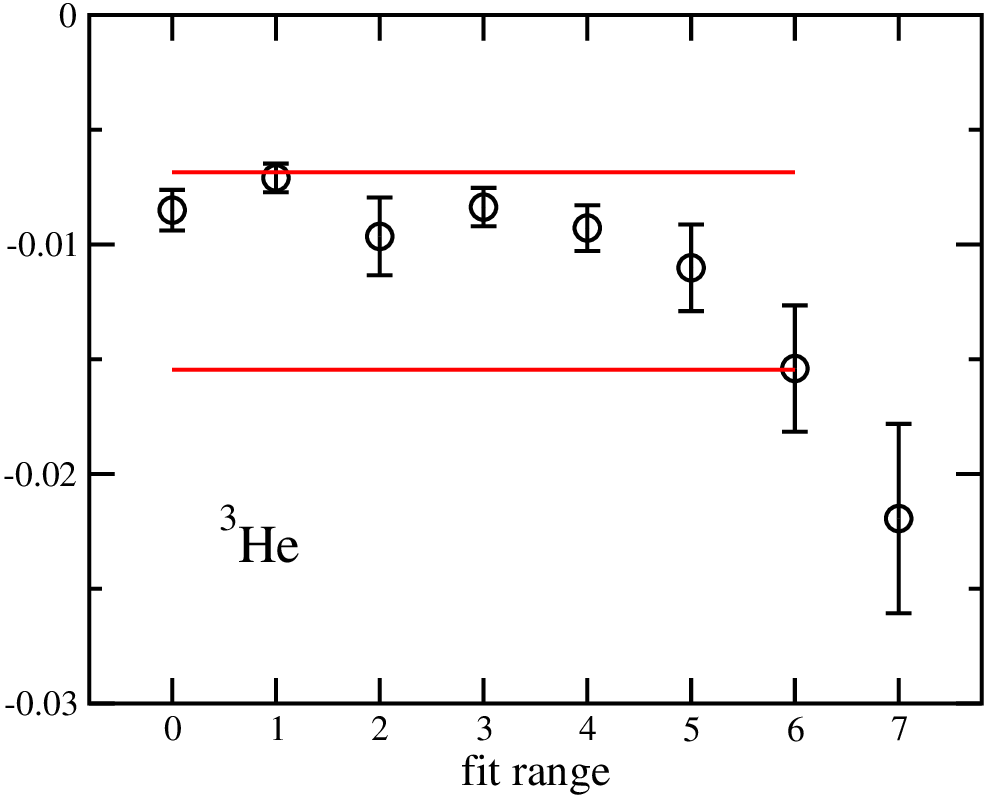}
\includegraphics*[angle=0,width=0.6\textwidth]{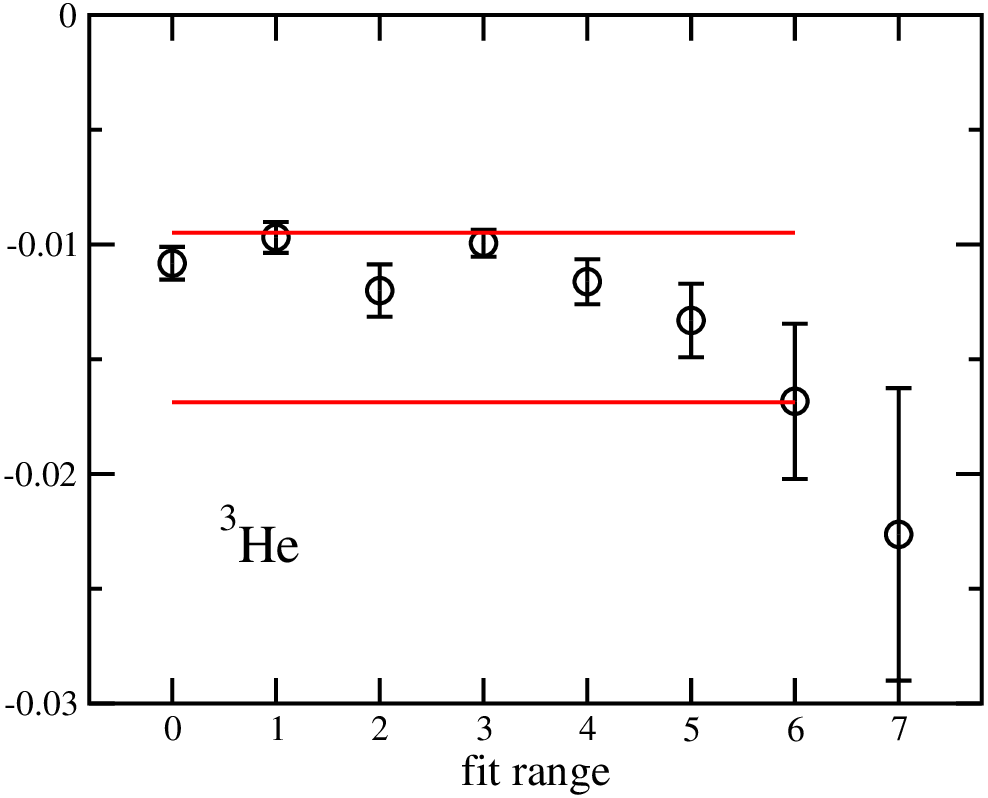}
\caption{
Same figure as Fig.~\ref{fig:fit_dep_h4}, but for $^3$He channel.
\label{fig:fit_dep_h3}
}
\end{figure}

Figure~\ref{fig:eff_R_h3} shows the effective energy shift 
$\Delta E^{\mathrm{eff}}_L$ in Eq.~(\ref{eq:eff_delE_L}) 
for the two volumes.
The signals are better than those in the 
$^4$He channel shown in Fig.~\ref{fig:eff_R_h4}.
A plateau is seen for  the smaller volume case,
while it is less clear in the region of $t=8$--12 for the larger volume case.
The energy shift $\Delta E_L$ in Eq.~(\ref{eq:delE_L}) is
determined by an exponential fit to $R(t)$ in Eq.~(\ref{eq:R})
with the fit range of $t=8$--12 and 9--12 for the
smaller and larger volumes, respectively.
The systematic error of $\Delta E_L$ is estimated in the same way
as for the $^4$He case as described in the above subsection.
The fit results with the statistical and systematic errors are
shown in Fig.~\ref{fig:eff_R_h3} and Table~\ref{tab:dE_He}.
The explanations for the solid and dashed lines are given in the previous
subsection.
Figure~\ref{fig:fit_dep_h3} shows
how we estimate the systematic error.
The relative difference of the extra fit result with the slid fit 
range by $+3$, as explained in the previous subsection, 
from the central fit result with the total error
is less than 1.6 $\sigma$.

\begin{figure}[!t]
\includegraphics*[angle=0,width=0.6\textwidth]{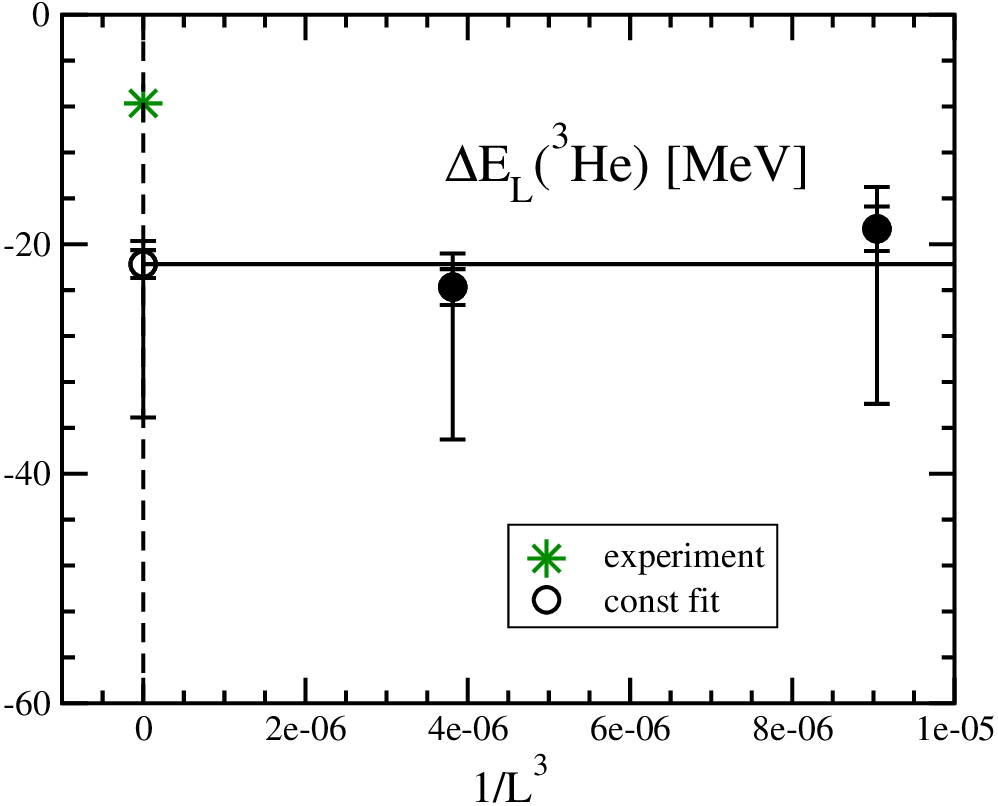}
\caption{
Same figure as Fig.~\ref{fig:dE_h4}, but for $^3$He channel.
\label{fig:dE_h3}
}
\end{figure}

A weak volume dependence of $\Delta E_L$ observed in
Fig.~\ref{fig:dE_h3} is similar to those in the previous results~\cite{Yamazaki:2009ua,Beane:2012vq,Yamazaki:2012hi}.
A constant fit of $\Delta E_L$ with only the statistical error 
gives a large value of $\chi^2/$d.o.f. = 4.1.
It agrees with the two data within
the total error as shown in Fig.~\ref{fig:dE_h3}, however. 
Thus, we take the constant fit result as the estimate
of the central value of $\Delta E_\infty$ in this calculation.

We estimate the systematic error
of $\Delta E_\infty$ in the same way to the $^4$He case.
We omit 18 constant fit results, however, with $\chi^2/$d.o.f. $ > 4.1$.
The extrapolated result of $\Delta E_\infty$ is clearly nonzero and negative
as presented in Fig.~\ref{fig:dE_h3}.
Thus the ground state is a bound state, corresponding to the $^3$He nucleus, 
with a binding energy of 
$-\Delta E_\infty = 21.7(1.2)(^{+13}_{-1.6})$ MeV, where
the first and second errors are statistical and systematic, respectively.

\begin{figure}[!t]
\includegraphics*[angle=0,width=0.6\textwidth]{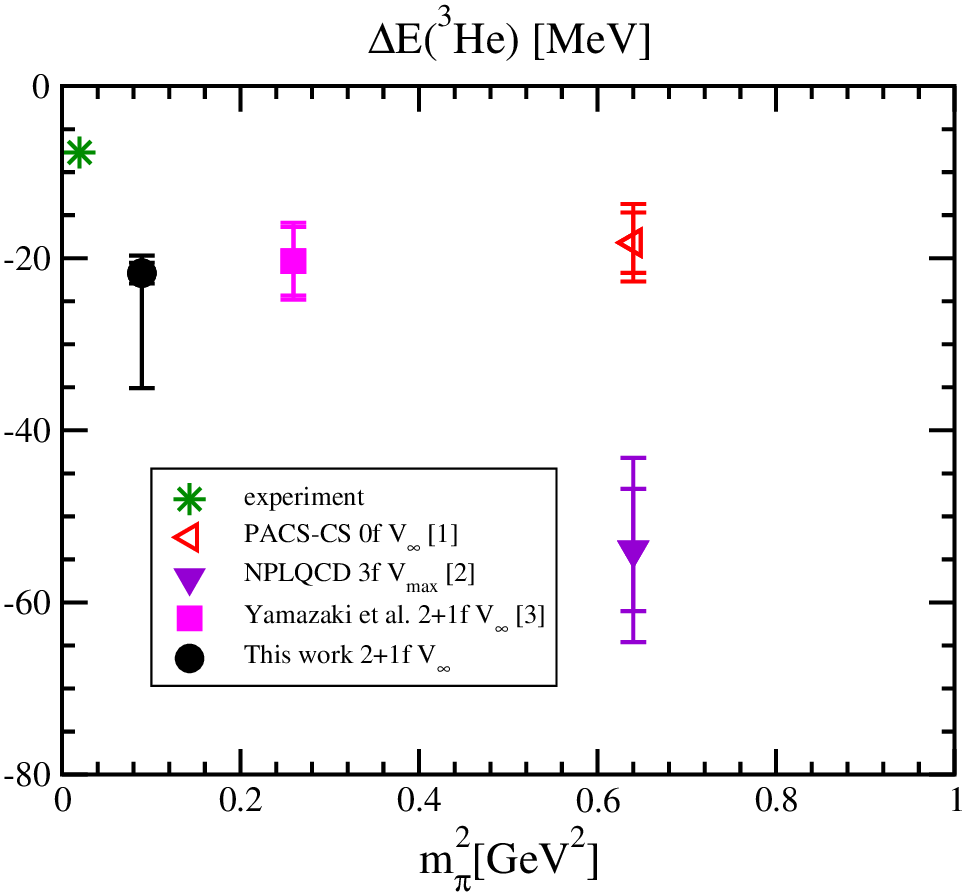}
\caption{
Same figure as Fig.~\ref{fig:mdep_4He}, but for $^3$He channel.
\label{fig:mdep_3He}
}
\end{figure}

The quark mass dependence of the energy shift is plotted in Fig.~\ref{fig:mdep_3He}. 
Our present result together with our two previous calculations~\cite{Yamazaki:2009ua,Yamazaki:2012hi} show very small dependence, 
while NPLQCD reported a much deeper bound state~\cite{Beane:2012vq}.
All  lattice results in the figure have the binding energy larger than  
the experimental value 7.72 MeV.

\subsection{Two-nucleon channels}
\label{sec:2N}

\begin{figure}[!t]
\includegraphics*[angle=0,width=0.6\textwidth]{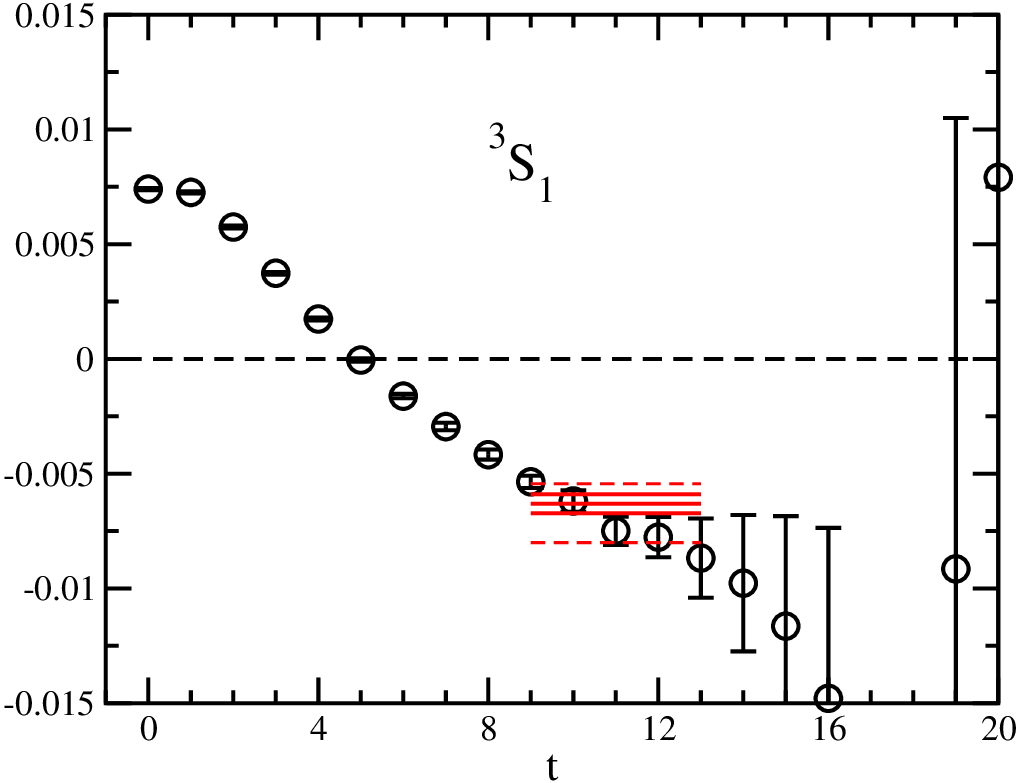}
\includegraphics*[angle=0,width=0.6\textwidth]{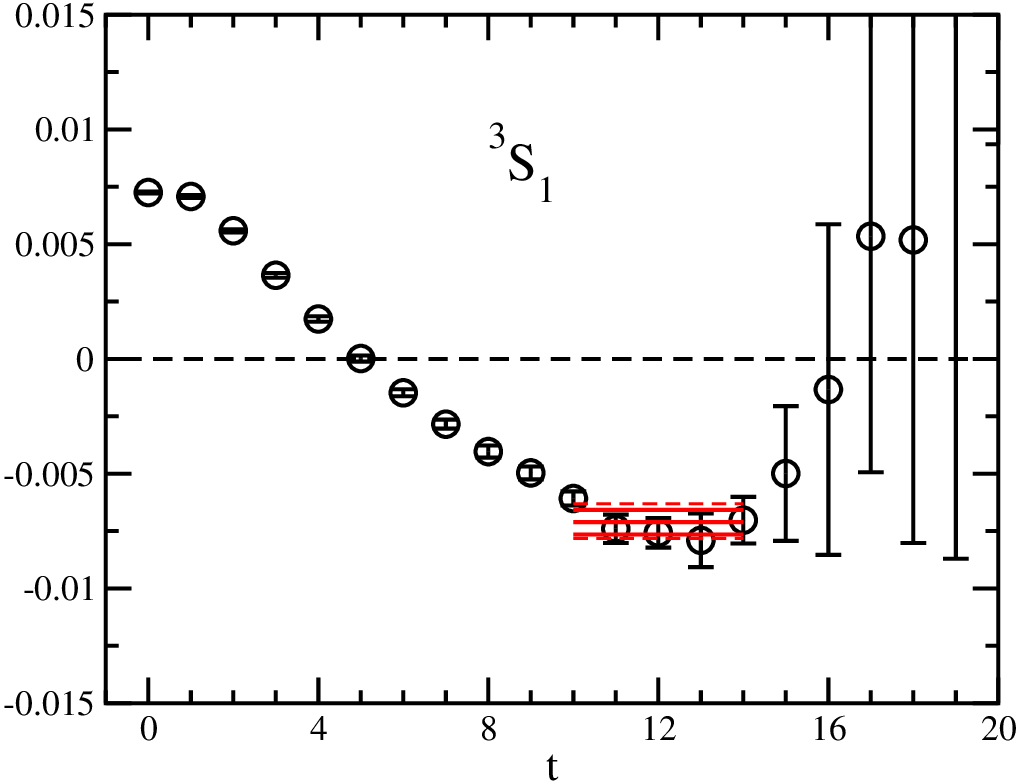}
\caption{
Same figure as Fig.~\ref{fig:eff_R_h4}, but for $^3$S$_1$ $NN$ channel.
\label{fig:eff_R_3S1}
}
\end{figure}

\begin{figure}[!t]
\includegraphics*[angle=0,width=0.6\textwidth]{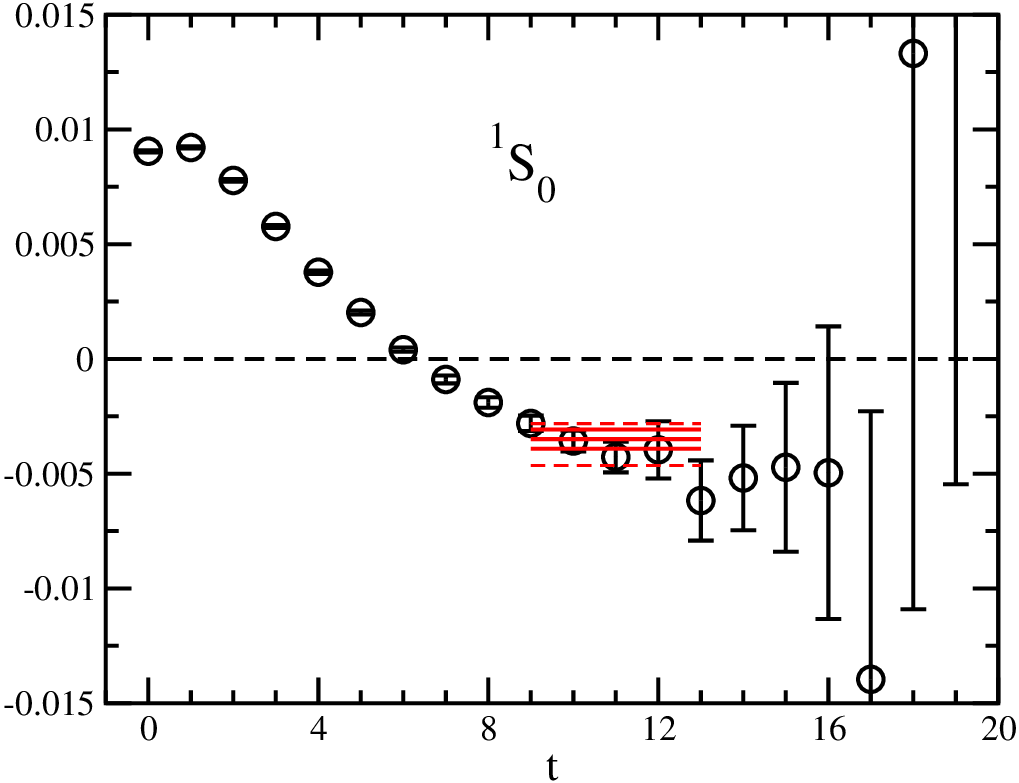}
\includegraphics*[angle=0,width=0.6\textwidth]{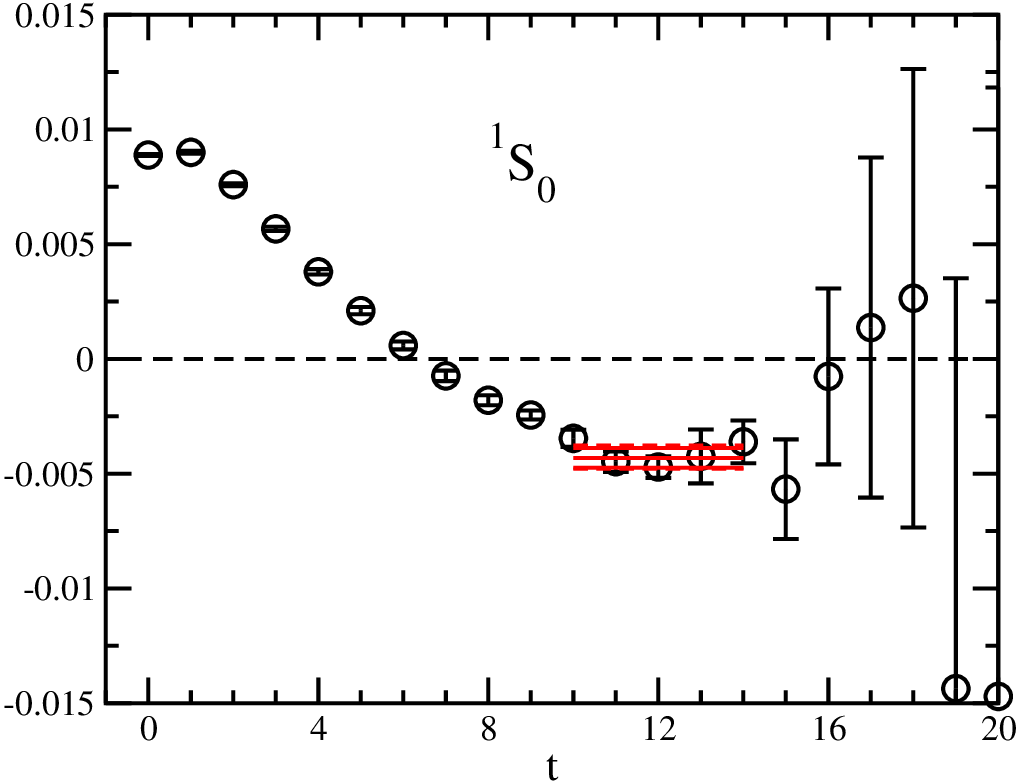}
\caption{
Same figure as Fig.~\ref{fig:eff_R_h4}, but for $^1$S$_0$ $NN$ channel.
\label{fig:eff_R_1S0}
}
\end{figure}

\begin{figure}[!t]
\includegraphics*[angle=0,width=0.6\textwidth]{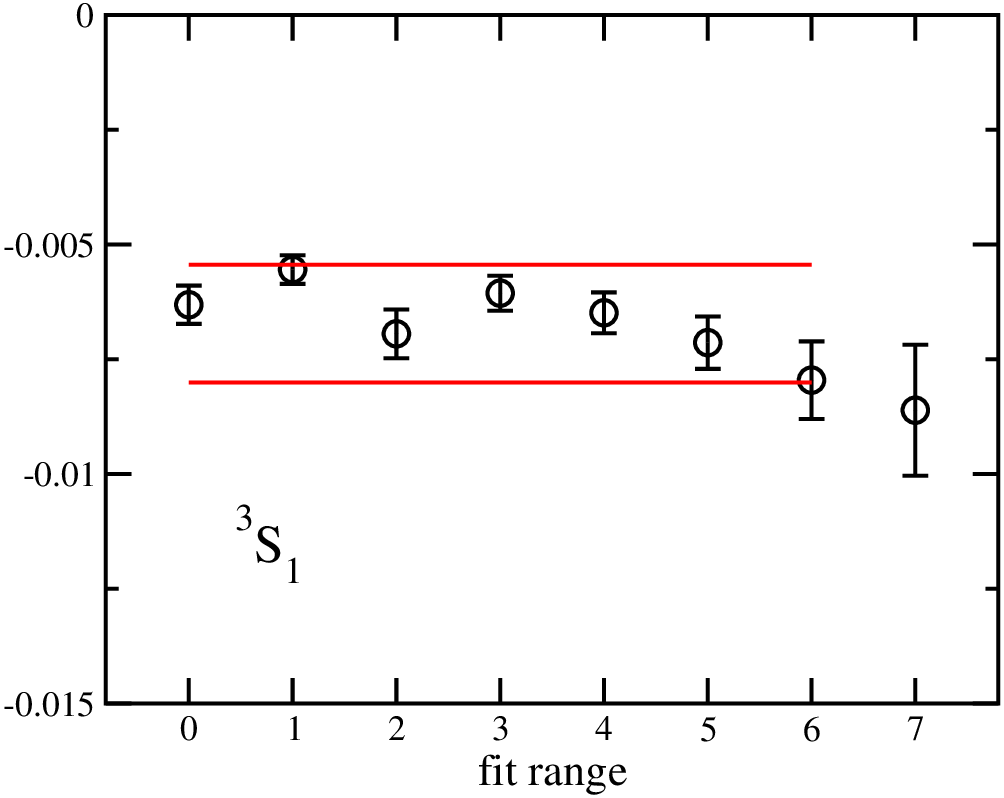}
\includegraphics*[angle=0,width=0.6\textwidth]{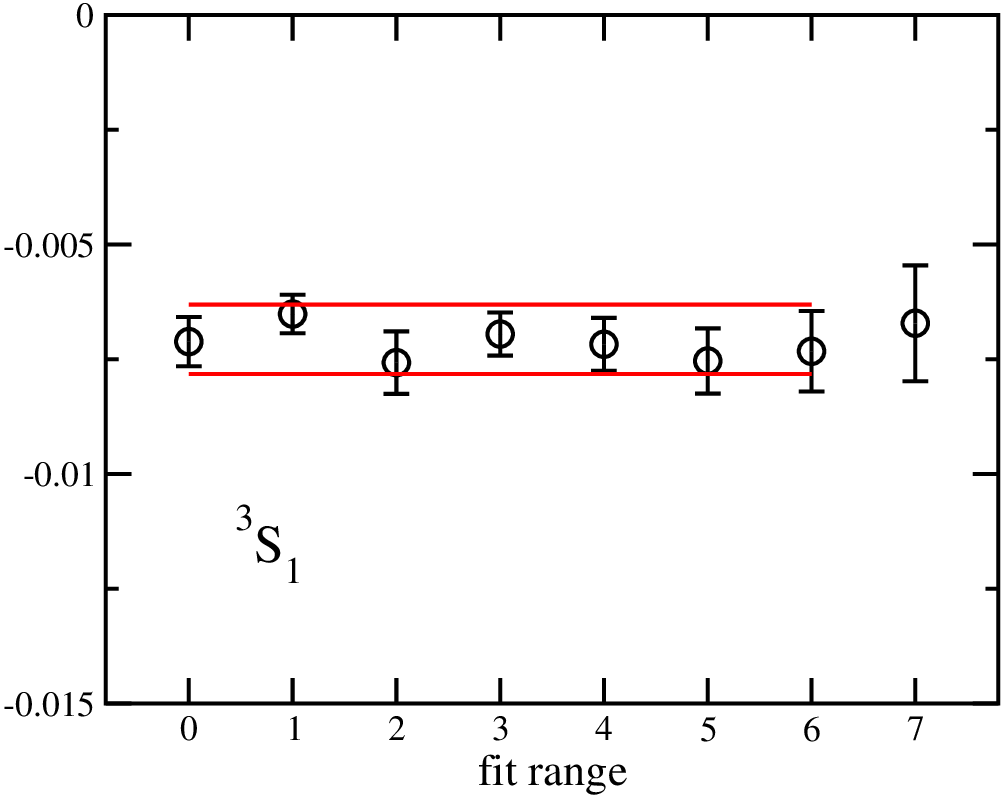}
\caption{
Same figure as Fig.~\ref{fig:fit_dep_h4}, but for $^3$S$_1$ $NN$ channel.
\label{fig:fit_dep_3S1}
}
\end{figure}

\begin{figure}[!t]
\includegraphics*[angle=0,width=0.6\textwidth]{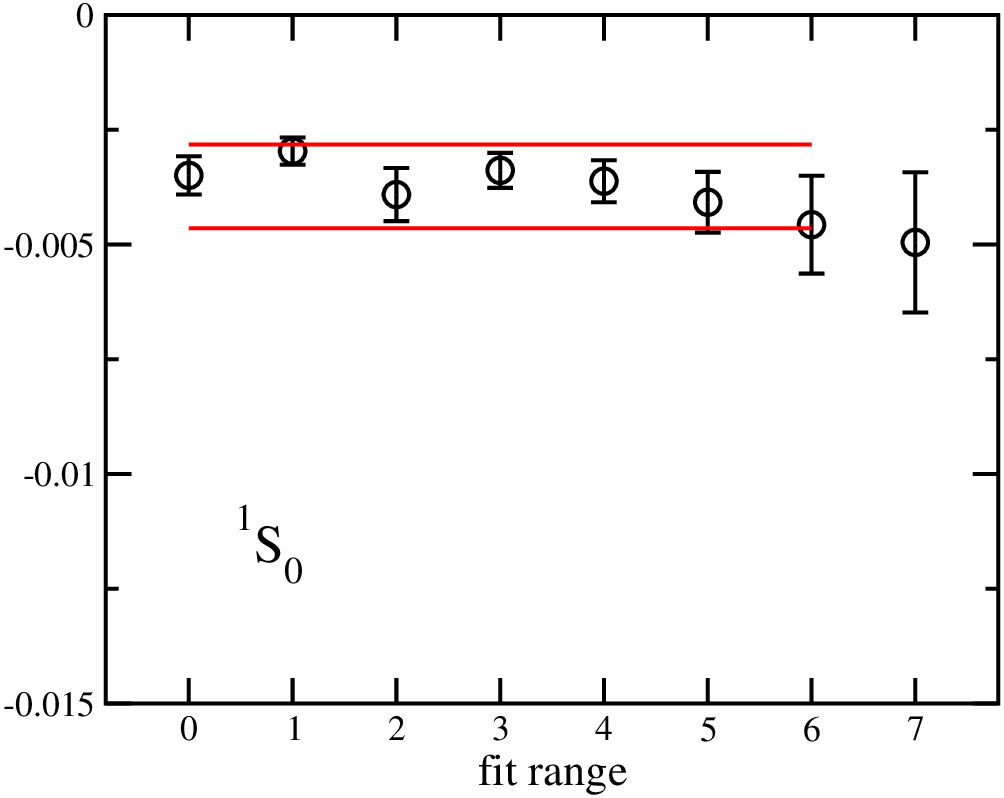}
\includegraphics*[angle=0,width=0.6\textwidth]{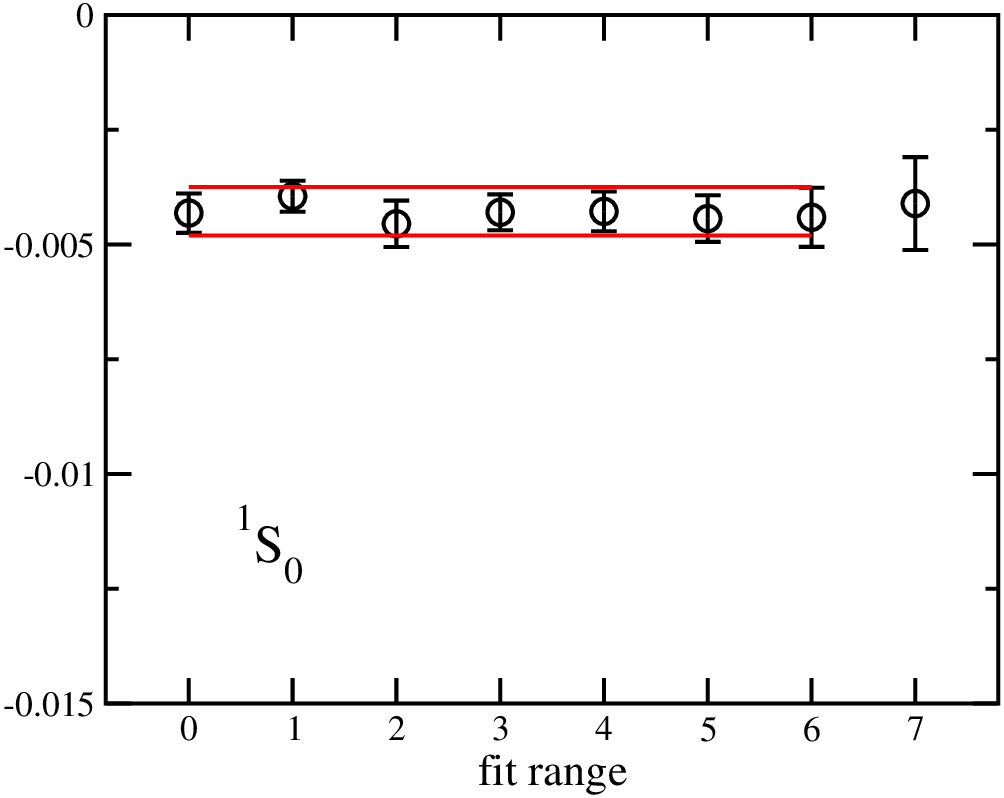}
\caption{
Same figure as Fig.~\ref{fig:fit_dep_h4}, but for $^1$S$_0$ $NN$ channel.
\label{fig:fit_dep_1S0}
}
\end{figure}

We present $\Delta E_L^{\mathrm{eff}}$ in Eq.~(\ref{eq:eff_delE_L})
for the $^3$S$_1$ and $^1$S$_0$ channels
in Figs.~\ref{fig:eff_R_3S1} and~\ref{fig:eff_R_1S0}, respectively.
The signals are clean up to $t \approx 14$, 
but statistical fluctuations spoil the signals in the larger time region.
The values of $|\Delta E_L^{\mathrm{eff}}|$ in the $^1$S$_0$ channel
are smaller than those in the $^3$S$_1$ channel.
A similar trend was seen in the 
previous studies~\cite{Yamazaki:2011nd,Beane:2012vq,Yamazaki:2012hi}.
We observe a clear plateau with a negative energy shift for $9 \simlt t \simlt 14$,
although of a less quality for the $^3$S$_1$ channel for the smaller volume.
We determine $\Delta E_L$ by an exponential fit to $R(t)$ of
Eq.~(\ref{eq:R}) with the fixed fit range of $t=9$--13 for 
the $^3$S$_1$ channel, and with $t=10-14$ for $^1$S$_0$.
The fit results are presented in 
Figs.~\ref{fig:eff_R_3S1} and~\ref{fig:eff_R_1S0},
and are summarized in Table~\ref{tab:dE_NN}.
The systematic error estimations for the $^3$S$_1$ and $^1$S$_0$ channels
using the results with the several fit ranges,
as in the $^4$He and $^3$He cases,
are presented in Figs.~\ref{fig:fit_dep_3S1} 
and~\ref{fig:fit_dep_1S0}, respectively.

\begin{figure}[!t]
\includegraphics*[angle=0,width=0.6\textwidth]{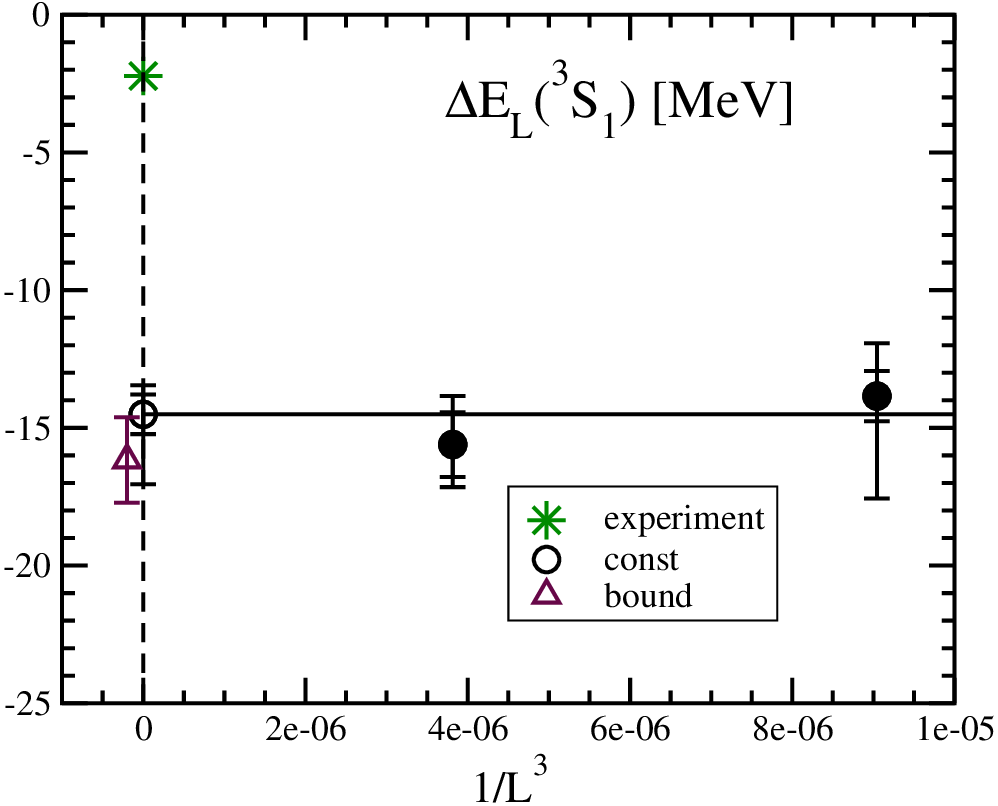}
\caption{
Same figure as Fig.~\ref{fig:dE_h4}, but for $^3$S$_1$ $NN$ channel.
Fit result using finite volume dependence of two-particle binding energy
Eq.(\ref{eq:FVE_BND}) is also plotted.
\label{fig:dE_3S1}
}
\end{figure}

\begin{figure}[!t]
\includegraphics*[angle=0,width=0.6\textwidth]{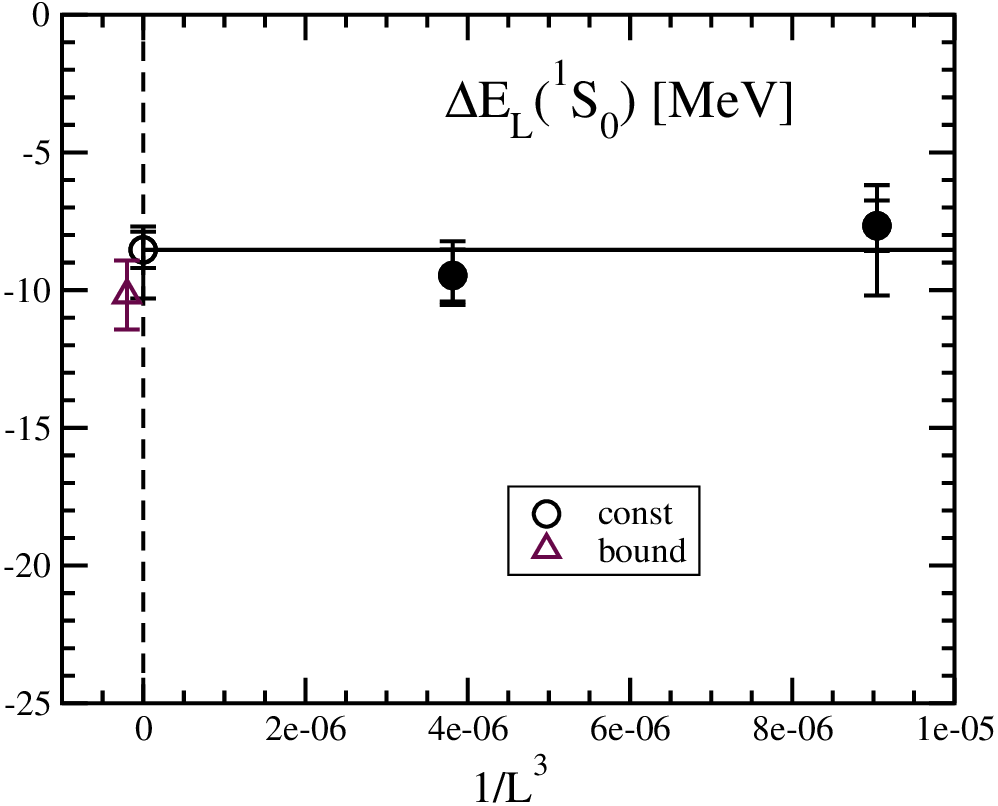}
\caption{
Same figure as Fig.~\ref{fig:dE_3S1}, but for $^1$S$_0$ $NN$ channel.
\label{fig:dE_1S0}
}
\end{figure}

The volume dependence of the energy shift in the two channels is shown in 
Figs.~\ref{fig:dE_3S1} and~\ref{fig:dE_1S0}.
In both channels, the volume dependences are mild,
so that the data can be reasonably fitted by a constant.
The results for the constant fit are non-zero and negative.
This indicates that the ground states in the two channels
are bound states.
The same conclusion is 
also obtained by a fit including finite volume effects of the 
two-particle bound state~\cite{Beane:2003da,Sasaki:2006jn},
\begin{equation}
\Delta E_L = -\frac{\gamma^2}{m_N}\left\{
1 + \frac{C_\gamma}{\gamma L} \sum^{\hspace{6mm}\prime}_{\vec{n}}
\frac{\exp(-\gamma L \sqrt{\vec{n}^2})}{\sqrt{\vec{n}^2}}
\right\},
\label{eq:FVE_BND}
\end{equation}
where $\gamma$ and $C_\gamma$ are free parameters, 
$\vec{n}$ is a three-dimensional integer vector,
and $\sum^\prime_{\vec{n}}$ denotes the summation without $|\vec{n}|=0$.
In the fit, we use the weighted average value of $m_N$ with the two volume data.
In the above equation, it is assumed that
\begin{equation}
-\Delta E_\infty = \frac{\gamma^2}{m_N} \approx
2 m_N - 2\sqrt{m_N^2 - \gamma^2}.
\end{equation}
Note that the degrees of freedom are zero
in the fit with Eq.(\ref{eq:FVE_BND}).
The fit result is presented in each figure at $1/L^3=0$.
We take the constant fit as the central value of the binding
energy $-\Delta E_\infty$,
and estimate the systematic error in the same way as in 
other channels.
In the systematic error estimation, we include 
the fit result using Eq.~(\ref{eq:FVE_BND}),
while we exclude 9 and 8 constant fit results in the $^3$S$_1$ and $^1$S$_0$
channels, respectively, which yield $\chi^2/$d.o.f. $ > 3$.
The results for the binding energy are
$-\Delta E_\infty = 14.5(0.7)(^{+2.4}_{-0.8})$ MeV 
for the $^3$S$_1$ channel and
$8.5(0.7)(^{+1.6}_{-0.5})$ MeV for $^1$S$_0$
with the first and second errors being the statistical and systematic,
which are also summarized in Table~\ref{tab:dE_NN}.

\begin{figure}[!t]
\includegraphics*[angle=0,width=0.6\textwidth]{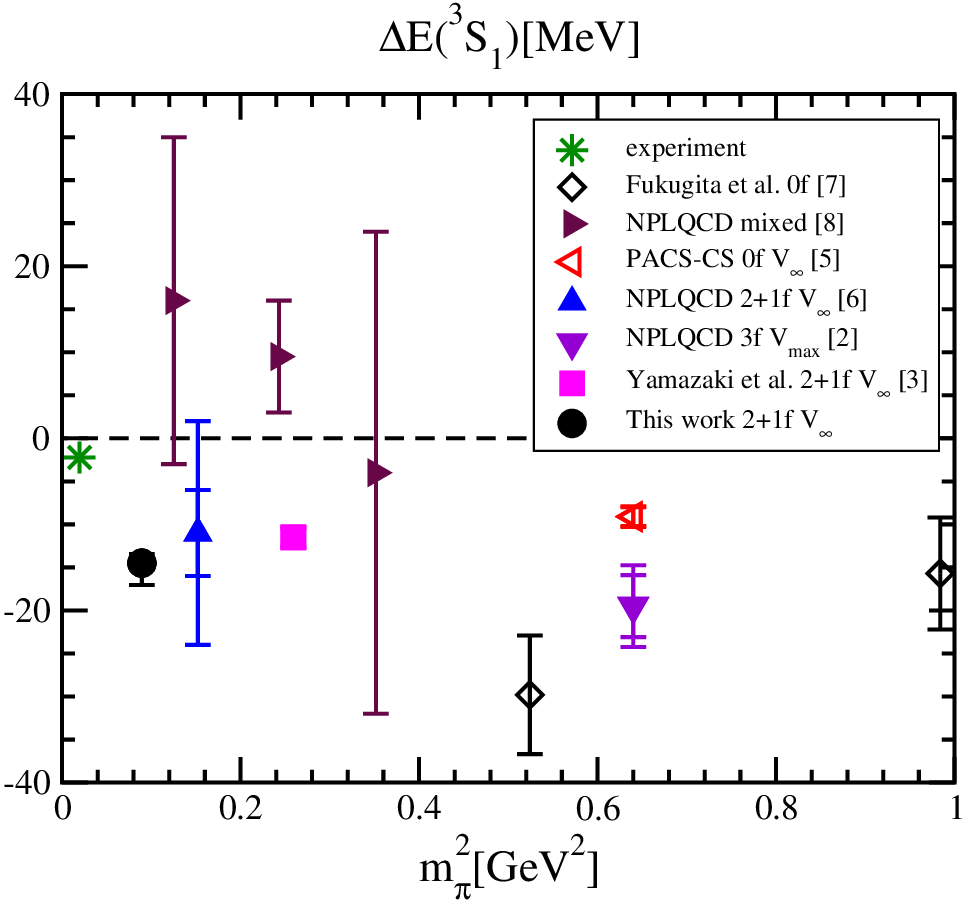}
\caption{
Same figure as Fig.~\ref{fig:mdep_4He}, but for $^3$S$_1$ $NN$ channel.
Open and closed symbols denote the quenched~\cite{Fukugita:1994ve,Yamazaki:2011nd} 
and full QCD~\cite{Beane:2006mx,Beane:2011iw,Beane:2012vq,Yamazaki:2012hi}
results, respectively. 
The results of Refs.~\cite{Yamazaki:2011nd,Beane:2011iw,Beane:2012vq,Yamazaki:2012hi} 
and this work are the ones in the infinite volume limit.
\label{fig:mdep_3S1}
}
\end{figure}

\begin{figure}[!t]
\includegraphics*[angle=0,width=0.6\textwidth]{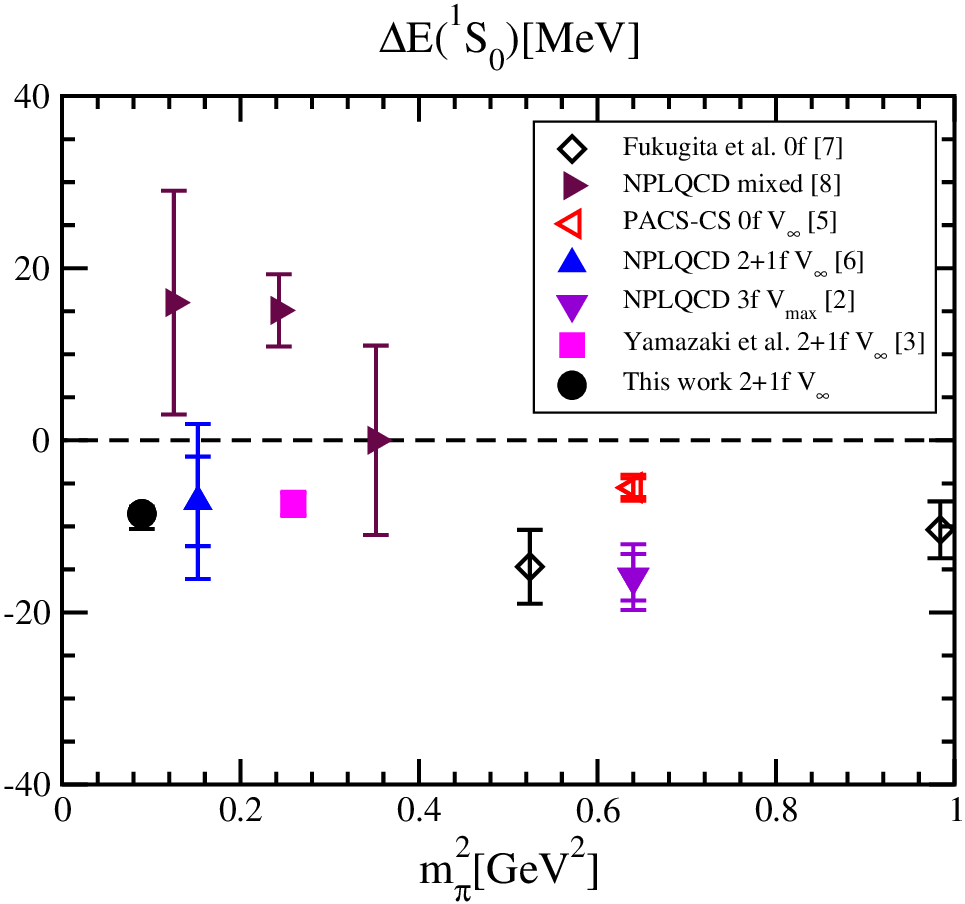}
\caption{
Same figure as Fig.~\ref{fig:mdep_3S1}, but for $^1$S$_0$ $NN$ channel.
\label{fig:mdep_1S0}
}
\end{figure}

In Figs.~\ref{fig:mdep_3S1} and~\ref{fig:mdep_1S0}, 
the results for $\Delta E_\infty$ in the present work are compared with those of
the previous studies~\cite{Fukugita:1994ve,Beane:2006mx,Yamazaki:2011nd,Beane:2011iw,Beane:2012vq,Yamazaki:2012hi}
as a function of $m_\pi^2$.
Almost all results report negative values,
except for those of Ref.~\cite{Beane:2006mx} with large errors.
The earlier calculations~\cite{Fukugita:1994ve,Beane:2006mx}
did not investigate the volume dependence of $\Delta E_L$. 
More recent studies~\cite{Yamazaki:2011nd,Beane:2011iw,Beane:2012vq,Yamazaki:2012hi} 
examined the dependence and estimated
the infinite volume value through extrapolations~\cite{Yamazaki:2011nd,Beane:2011iw,Yamazaki:2012hi} or checked that there is no significant volume dependence
of $\Delta E_L$~\cite{Beane:2012vq}.
All the recent results suggest that
the ground states in both the channels are bound states.  
One exception is Ref.\cite{Beane:2011iw} where the conclusion is not clear due to 
large errors.

While lattice results are mutually qualitatively consistent, 
they differ from experiment in more than one aspects. 
For the $^3$S$_1$ channel,
the binding energy $-\Delta E_\infty$ found in the lattice calculations~\cite{Yamazaki:2011nd,Beane:2011iw,Beane:2012vq,Yamazaki:2012hi} 
is a factor five to ten times larger than the experimental value.
Furthermore, we observe no tendency in the binding
energy to approach the experimental value, at least over the pion mass range  $m_\pi = 0.3$--0.51 GeV.  
For the $^1$S$_0$ channel,  
the bound state found in the lattice calculations is absent 
in experiment. 
Furthermore, similarly to the $^3$S$_1$ channel,
the binding energy is almost flat in $m_\pi^2$ 
in the interval $m_\pi = 0.30$--0.51 GeV.
It is not clear whether the bound state observed in the lattice
calculation becomes unbound toward the physical $m_\pi$.

\section{Conclusion and discussion}
\label{sec:conclusion}

We have extended our previous nuclei calculation in 2+1 flavor QCD
at $m_\pi = 0.51$ GeV~\cite{Yamazaki:2012hi}
to the lighter quark mass corresponding to
$m_\pi=0.30$ GeV and $m_N=1.05$ GeV.
In order to suppress an exponential increase of statistical errors 
at smaller $m_\pi$,
we have carried out a much larger number of measurements by a factor twelve and five 
for the case of the spatial extent of 4.3 fm ($48^3$) and 5.8 fm ($64^3$), respectively,
compared to those for the $m_\pi = 0.51$ GeV case with  the same volumes.
We have found that in all channels we have studied, $^4$He, $^3$He,
and two-nucleon $^3$S$_1$ and $^1$S$_0$, the ground state is a bound state 
by investigating the volume dependence of energy shift $\Delta E_L$. 
The binding energies estimated for the infinite volume are as follows:
\begin{equation}
-\Delta E_{\infty} = \left\{
\begin{array}{ccl}
47(7)(^{+20}_{-11}) & \mathrm{MeV} & \mathrm{for}\ ^4\mathrm{He},\\
21.7(1.2)(^{+13}_{-1.6}) & \mathrm{MeV} & \mathrm{for}\ ^3\mathrm{He},\\
14.5(0.7)(^{+2.4}_{-0.8}) & \mathrm{MeV} & \mathrm{for}\ ^3\mathrm{S}_1,\\
8.5(0.7)(^{+1.6}_{-0.5}) & \mathrm{MeV} & \mathrm{for}\ ^1\mathrm{S}_0.\\
\end{array}
\right.
\end{equation}
These values differ little from those obtained at $m_\pi = 0.51$ GeV~\cite{Yamazaki:2012hi}.
The largest relative difference occurs for the $^3$S$_1$ channel, which is only a 1.9 $\sigma$ 
effect if we use the total error adding the statistical and systematic ones by quadrature.
Therefore, our conclusions at $m_\pi=0.30$~GeV are  similar to those in 
Ref.~\cite{Yamazaki:2012hi} for $m_\pi=0.51$~GeV:
the binding energy of the $^4$He nucleus is comparable with the experimental 
value, while the $^3$He nucleus and the deuteron are 
about three and seven times larger than the experimental values,
respectively, and a bound dineutron is observed in the $^1$S$_0$ channel.

The differences we observe from experiment may arise from various sources, either computational or physical in origin.  
Statistical errors are fairly large in the calculations even for light nuclei.  While the negative value of the energy shift is certain in all channels we looked at, better statistics and improved techniques will be welcome to better control the extraction of the energy shift for each volume and the infinite volume extrapolations.    

The quark mass is heavier than experiment in all calculations to date.
The binding in the $^1$S$_0$ channel is shallower than the  $^3$S$_1$ so that the former bound state may become unbound as $m_\pi$ decreases toward the physical value.  
This can only be verified by calculations of the nuclear binding energy 
at smaller quark masses.  

It is also possible that finite lattice spacing effect is rather subtle.  
The short distance repulsion, in the language of nuclear potential, 
is possibly affected more by such effects than the long distance attraction, 
so that finite lattice spacing effects may push out multi-nucleon wave function,
and then the ground state would become a scattering state,
for lattice spacings smaller than some value.  

Another possible source of systematic error is excited state contaminations
in the calculation.
We have assumed that
the nucleon and nucleus correlation functions are dominated by the ground state
in the large $t$ region, where the plateau of 
$\Delta E_L^{\mathrm{eff}}$ in Eq.~(\ref{eq:eff_delE_L}) appears.
While we have tuned the smearing parameter of the quark field to increase the 
overlap of the nucleus operator to its ground state,
from the current data 
we cannot completely exclude the possibility that it is not sufficient to 
suppress the contaminations.
To investigate the size of possible contaminations,
we might try analyses with the variational method~\cite{Luscher:1990ck} 
using correlation function matrices.

For now, however,  
we think that a calculation at the physical point, 
keeping the lattice spacing, is the next step.

\section*{Acknowledgments}
Numerical calculations for the present work have been carried out
on the FX10 cluster system at Information Technology Center
of the University of Tokyo, 
on the T2K-Tsukuba cluster system and HA-PACS system and COMA system
at University of Tsukuba,
and on K computer at RIKEN Advanced Institute for Computational Science.
We thank the colleagues in the PACS-CS Collaboration for helpful
discussions and providing us the code used in this work.
This work is supported in part by Grants-in-Aid for Scientific Research
from the Ministry of Education, Culture, Sports, Science and Technology 
(Nos. 22244018, 25800138) and 
Grants-in-Aid of the Japanese Ministry for Scientific Research on Innovative 
Areas (Nos. 20105002, 21105501, 23105708).

\bibliography{./paper_0.3}

\begin{thebibliography}{30}
\expandafter\ifx\csname natexlab\endcsname\relax\def\natexlab#1{#1}\fi
\expandafter\ifx\csname bibnamefont\endcsname\relax
  \def\bibnamefont#1{#1}\fi
\expandafter\ifx\csname bibfnamefont\endcsname\relax
  \def\bibfnamefont#1{#1}\fi
\expandafter\ifx\csname citenamefont\endcsname\relax
  \def\citenamefont#1{#1}\fi
\expandafter\ifx\csname url\endcsname\relax
  \def\url#1{\texttt{#1}}\fi
\expandafter\ifx\csname urlprefix\endcsname\relax\def\urlprefix{URL }\fi
\providecommand{\bibinfo}[2]{#2}
\providecommand{\eprint}[2][]{\url{#2}}

\bibitem[{\citenamefont{Yamazaki et~al.}(2010)\citenamefont{Yamazaki,
  Kuramashi, and Ukawa}}]{Yamazaki:2009ua}
\bibinfo{author}{\bibfnamefont{T.}~\bibnamefont{Yamazaki}},
  \bibinfo{author}{\bibfnamefont{Y.}~\bibnamefont{Kuramashi}},
  \bibnamefont{and} \bibinfo{author}{\bibfnamefont{A.}~\bibnamefont{Ukawa}}
  (\bibinfo{collaboration}{PACS-CS Collaboration}),
  \bibinfo{journal}{Phys.Rev.} \textbf{\bibinfo{volume}{D81}},
  \bibinfo{pages}{111504} (\bibinfo{year}{2010}), \eprint{0912.1383}.

\bibitem[{\citenamefont{Beane et~al.}(2013)\citenamefont{Beane, Chang, Cohen,
  Detmold, Lin et~al.}}]{Beane:2012vq}
\bibinfo{author}{\bibfnamefont{S.}~\bibnamefont{Beane}},
  \bibinfo{author}{\bibfnamefont{E.}~\bibnamefont{Chang}},
  \bibinfo{author}{\bibfnamefont{S.}~\bibnamefont{Cohen}},
  \bibinfo{author}{\bibfnamefont{W.}~\bibnamefont{Detmold}},
  \bibinfo{author}{\bibfnamefont{H.}~\bibnamefont{Lin}}, \bibnamefont{et~al.},
  \bibinfo{journal}{Phys.Rev.} \textbf{\bibinfo{volume}{D87}},
  \bibinfo{pages}{034506} (\bibinfo{year}{2013}), \eprint{1206.5219}.

\bibitem[{\citenamefont{Yamazaki et~al.}(2012)\citenamefont{Yamazaki, Ishikawa,
  Kuramashi, and Ukawa}}]{Yamazaki:2012hi}
\bibinfo{author}{\bibfnamefont{T.}~\bibnamefont{Yamazaki}},
  \bibinfo{author}{\bibfnamefont{K.-i.} \bibnamefont{Ishikawa}},
  \bibinfo{author}{\bibfnamefont{Y.}~\bibnamefont{Kuramashi}},
  \bibnamefont{and} \bibinfo{author}{\bibfnamefont{A.}~\bibnamefont{Ukawa}},
  \bibinfo{journal}{Phys.Rev.} \textbf{\bibinfo{volume}{D86}},
  \bibinfo{pages}{074514} (\bibinfo{year}{2012}), \eprint{1207.4277}.

\bibitem[{\citenamefont{Inoue et~al.}(2012)}]{Inoue:2011ai}
\bibinfo{author}{\bibfnamefont{T.}~\bibnamefont{Inoue}} \bibnamefont{et~al.}
  (\bibinfo{collaboration}{HAL QCD Collaboration}), \bibinfo{journal}{Nucl.
  Phys.} \textbf{\bibinfo{volume}{A881}}, \bibinfo{pages}{28}
  (\bibinfo{year}{2012}), \eprint{1112.5926}.

\bibitem[{\citenamefont{Yamazaki et~al.}(2011)\citenamefont{Yamazaki,
  Kuramashi, and Ukawa}}]{Yamazaki:2011nd}
\bibinfo{author}{\bibfnamefont{T.}~\bibnamefont{Yamazaki}},
  \bibinfo{author}{\bibfnamefont{Y.}~\bibnamefont{Kuramashi}},
  \bibnamefont{and} \bibinfo{author}{\bibfnamefont{A.}~\bibnamefont{Ukawa}},
  \bibinfo{journal}{Phys. Rev.} \textbf{\bibinfo{volume}{D84}},
  \bibinfo{pages}{054506} (\bibinfo{year}{2011}), \eprint{1105.1418}.

\bibitem[{\citenamefont{Beane et~al.}(2012)}]{Beane:2011iw}
\bibinfo{author}{\bibfnamefont{S.}~\bibnamefont{Beane}} \bibnamefont{et~al.}
  (\bibinfo{collaboration}{NPLQCD Collaboration}), \bibinfo{journal}{Phys.
  Rev.} \textbf{\bibinfo{volume}{D85}}, \bibinfo{pages}{054511}
  (\bibinfo{year}{2012}), \eprint{1109.2889}.

\bibitem[{\citenamefont{Fukugita et~al.}(1995)\citenamefont{Fukugita,
  Kuramashi, Okawa, Mino, and Ukawa}}]{Fukugita:1994ve}
\bibinfo{author}{\bibfnamefont{M.}~\bibnamefont{Fukugita}},
  \bibinfo{author}{\bibfnamefont{Y.}~\bibnamefont{Kuramashi}},
  \bibinfo{author}{\bibfnamefont{M.}~\bibnamefont{Okawa}},
  \bibinfo{author}{\bibfnamefont{H.}~\bibnamefont{Mino}}, \bibnamefont{and}
  \bibinfo{author}{\bibfnamefont{A.}~\bibnamefont{Ukawa}},
  \bibinfo{journal}{Phys. Rev.} \textbf{\bibinfo{volume}{D52}},
  \bibinfo{pages}{3003} (\bibinfo{year}{1995}), \eprint{hep-lat/9501024}.

\bibitem[{\citenamefont{Beane et~al.}(2006)\citenamefont{Beane, Bedaque,
  Orginos, and Savage}}]{Beane:2006mx}
\bibinfo{author}{\bibfnamefont{S.~R.} \bibnamefont{Beane}},
  \bibinfo{author}{\bibfnamefont{P.~F.} \bibnamefont{Bedaque}},
  \bibinfo{author}{\bibfnamefont{K.}~\bibnamefont{Orginos}}, \bibnamefont{and}
  \bibinfo{author}{\bibfnamefont{M.~J.} \bibnamefont{Savage}},
  \bibinfo{journal}{Phys. Rev. Lett.} \textbf{\bibinfo{volume}{97}},
  \bibinfo{pages}{012001} (\bibinfo{year}{2006}), \eprint{hep-lat/0602010}.

\bibitem[{\citenamefont{Ishii}(2013)}]{Ishii:2013ira}
\bibinfo{author}{\bibfnamefont{N.}~\bibnamefont{Ishii}}
  (\bibinfo{collaboration}{HAL QCD}), \bibinfo{journal}{PoS}
  \textbf{\bibinfo{volume}{CD12}}, \bibinfo{pages}{025} (\bibinfo{year}{2013}).

\bibitem[{\citenamefont{Iwasaki}(2011)}]{Iwasaki:2011jk}
\bibinfo{author}{\bibfnamefont{Y.}~\bibnamefont{Iwasaki}}
  (\bibinfo{year}{2011}), \bibinfo{note}{{UTHEP-118}}, \eprint{arXiv:1111.7054
  [hep-lat]}.

\bibitem[{\citenamefont{Aoki et~al.}(2006)}]{Aoki:2005et}
\bibinfo{author}{\bibfnamefont{S.}~\bibnamefont{Aoki}} \bibnamefont{et~al.}
  (\bibinfo{collaboration}{CP-PACS/JLQCD Collaborations}),
  \bibinfo{journal}{Phys. Rev.} \textbf{\bibinfo{volume}{D73}},
  \bibinfo{pages}{034501} (\bibinfo{year}{2006}), \eprint{hep-lat/0508031}.

\bibitem[{\citenamefont{Aoki et~al.}(2010)}]{Aoki:2009ix}
\bibinfo{author}{\bibfnamefont{S.}~\bibnamefont{Aoki}} \bibnamefont{et~al.}
  (\bibinfo{collaboration}{PACS-CS Collaboration}), \bibinfo{journal}{Phys.
  Rev.} \textbf{\bibinfo{volume}{D81}}, \bibinfo{pages}{074503}
  (\bibinfo{year}{2010}), \eprint{0911.2561}.

\bibitem[{\citenamefont{Aoki et~al.}(2009)}]{Aoki:2008sm}
\bibinfo{author}{\bibfnamefont{S.}~\bibnamefont{Aoki}} \bibnamefont{et~al.}
  (\bibinfo{collaboration}{PACS-CS Collaboration}), \bibinfo{journal}{Phys.
  Rev.} \textbf{\bibinfo{volume}{D79}}, \bibinfo{pages}{034503}
  (\bibinfo{year}{2009}), \eprint{0807.1661}.

\bibitem[{\citenamefont{L{\"u}scher}(2003)}]{Luscher:2003vf}
\bibinfo{author}{\bibfnamefont{M.}~\bibnamefont{L{\"u}scher}},
  \bibinfo{journal}{JHEP} \textbf{\bibinfo{volume}{0305}}, \bibinfo{pages}{052}
  (\bibinfo{year}{2003}), \eprint{hep-lat/0304007}.

\bibitem[{\citenamefont{L{\"u}scher}(2005)}]{Luscher:2005rx}
\bibinfo{author}{\bibfnamefont{M.}~\bibnamefont{L{\"u}scher}},
  \bibinfo{journal}{Comput. Phys. Commun.} \textbf{\bibinfo{volume}{165}},
  \bibinfo{pages}{199} (\bibinfo{year}{2005}), \eprint{hep-lat/0409106}.

\bibitem[{\citenamefont{Hasenbusch}(2001)}]{Hasenbusch:2001ne}
\bibinfo{author}{\bibfnamefont{M.}~\bibnamefont{Hasenbusch}},
  \bibinfo{journal}{Phys.Lett.} \textbf{\bibinfo{volume}{B519}},
  \bibinfo{pages}{177} (\bibinfo{year}{2001}), \eprint{hep-lat/0107019}.

\bibitem[{\citenamefont{Ishikawa et~al.}(2006)}]{Ishikawa:2006pb}
\bibinfo{author}{\bibfnamefont{K.-I.} \bibnamefont{Ishikawa}}
  \bibnamefont{et~al.} (\bibinfo{collaboration}{PACS-CS Collaboration}),
  \bibinfo{journal}{PoS} \textbf{\bibinfo{volume}{LAT2006}},
  \bibinfo{pages}{027} (\bibinfo{year}{2006}), \eprint{hep-lat/0610037}.

\bibitem[{\citenamefont{Omelyan et~al.}(2003)\citenamefont{Omelyan, Mryglod,
  and Folk}}]{Omelyan:2003om}
\bibinfo{author}{\bibfnamefont{I.~P.} \bibnamefont{Omelyan}},
  \bibinfo{author}{\bibfnamefont{I.~M.} \bibnamefont{Mryglod}},
  \bibnamefont{and} \bibinfo{author}{\bibfnamefont{R.}~\bibnamefont{Folk}},
  \bibinfo{journal}{Comput. Phys. Commun.} \textbf{\bibinfo{volume}{151}},
  \bibinfo{pages}{272} (\bibinfo{year}{2003}).

\bibitem[{\citenamefont{Takaishi and de~Forcrand}(2006)}]{Takaishi:2005tz}
\bibinfo{author}{\bibfnamefont{T.}~\bibnamefont{Takaishi}} \bibnamefont{and}
  \bibinfo{author}{\bibfnamefont{P.}~\bibnamefont{de~Forcrand}},
  \bibinfo{journal}{Phys. Rev.} \textbf{\bibinfo{volume}{E73}},
  \bibinfo{pages}{036706} (\bibinfo{year}{2006}), \eprint{hep-lat/0505020}.

\bibitem[{\citenamefont{Detmold and Savage}(2010)}]{Detmold:2010au}
\bibinfo{author}{\bibfnamefont{W.}~\bibnamefont{Detmold}} \bibnamefont{and}
  \bibinfo{author}{\bibfnamefont{M.~J.} \bibnamefont{Savage}},
  \bibinfo{journal}{Phys. Rev.} \textbf{\bibinfo{volume}{D82}},
  \bibinfo{pages}{014511} (\bibinfo{year}{2010}), \eprint{1001.2768}.

\bibitem[{\citenamefont{Doi and Endres}(2013)}]{Doi:2012xd}
\bibinfo{author}{\bibfnamefont{T.}~\bibnamefont{Doi}} \bibnamefont{and}
  \bibinfo{author}{\bibfnamefont{M.~G.} \bibnamefont{Endres}},
  \bibinfo{journal}{Comput.Phys.Commun.} \textbf{\bibinfo{volume}{184}},
  \bibinfo{pages}{117} (\bibinfo{year}{2013}), \eprint{1205.0585}.

\bibitem[{\citenamefont{Detmold and Orginos}(2013)}]{Detmold:2012eu}
\bibinfo{author}{\bibfnamefont{W.}~\bibnamefont{Detmold}} \bibnamefont{and}
  \bibinfo{author}{\bibfnamefont{K.}~\bibnamefont{Orginos}},
  \bibinfo{journal}{Phys.Rev.} \textbf{\bibinfo{volume}{D87}},
  \bibinfo{pages}{114512} (\bibinfo{year}{2013}), \eprint{1207.1452}.

\bibitem[{\citenamefont{G{\"u}nther et~al.}(2013)\citenamefont{G{\"u}nther,
  T{\'o}th, and Varnhorst}}]{Gunther:2013xj}
\bibinfo{author}{\bibfnamefont{J.}~\bibnamefont{G{\"u}nther}},
  \bibinfo{author}{\bibfnamefont{B.~C.} \bibnamefont{T{\'o}th}},
  \bibnamefont{and}
  \bibinfo{author}{\bibfnamefont{L.}~\bibnamefont{Varnhorst}},
  \bibinfo{journal}{Phys.Rev.} \textbf{\bibinfo{volume}{D87}},
  \bibinfo{pages}{094513} (\bibinfo{year}{2013}), \eprint{1301.4895}.

\bibitem[{\citenamefont{L{\"u}scher}(1986)}]{Luscher:1986pf}
\bibinfo{author}{\bibfnamefont{M.}~\bibnamefont{L{\"u}scher}},
  \bibinfo{journal}{Commun. Math. Phys.} \textbf{\bibinfo{volume}{105}},
  \bibinfo{pages}{153} (\bibinfo{year}{1986}).

\bibitem[{\citenamefont{Beane et~al.}(2004)\citenamefont{Beane, Bedaque,
  Parreno, and Savage}}]{Beane:2003da}
\bibinfo{author}{\bibfnamefont{S.~R.} \bibnamefont{Beane}},
  \bibinfo{author}{\bibfnamefont{P.~F.} \bibnamefont{Bedaque}},
  \bibinfo{author}{\bibfnamefont{A.}~\bibnamefont{Parreno}}, \bibnamefont{and}
  \bibinfo{author}{\bibfnamefont{M.~J.} \bibnamefont{Savage}},
  \bibinfo{journal}{Phys. Lett.} \textbf{\bibinfo{volume}{B585}},
  \bibinfo{pages}{106} (\bibinfo{year}{2004}), \eprint{hep-lat/0312004}.

\bibitem[{\citenamefont{Sasaki and Yamazaki}(2006)}]{Sasaki:2006jn}
\bibinfo{author}{\bibfnamefont{S.}~\bibnamefont{Sasaki}} \bibnamefont{and}
  \bibinfo{author}{\bibfnamefont{T.}~\bibnamefont{Yamazaki}},
  \bibinfo{journal}{Phys. Rev.} \textbf{\bibinfo{volume}{D74}},
  \bibinfo{pages}{114507} (\bibinfo{year}{2006}), \eprint{hep-lat/0610081}.

\bibitem[{\citenamefont{Beane et~al.}(2007)\citenamefont{Beane, Detmold, and
  Savage}}]{Beane:2007qr}
\bibinfo{author}{\bibfnamefont{S.~R.} \bibnamefont{Beane}},
  \bibinfo{author}{\bibfnamefont{W.}~\bibnamefont{Detmold}}, \bibnamefont{and}
  \bibinfo{author}{\bibfnamefont{M.~J.} \bibnamefont{Savage}},
  \bibinfo{journal}{Phys. Rev.} \textbf{\bibinfo{volume}{D76}},
  \bibinfo{pages}{074507} (\bibinfo{year}{2007}), \eprint{0707.1670}.

\bibitem[{\citenamefont{Beam}(1967)}]{Beam:1967zz}
\bibinfo{author}{\bibfnamefont{J.~E.} \bibnamefont{Beam}},
  \bibinfo{journal}{Phys. Rev.} \textbf{\bibinfo{volume}{158}},
  \bibinfo{pages}{907} (\bibinfo{year}{1967}).

\bibitem[{\citenamefont{Bolsterli and Jezak}(1964)}]{Bolsterli:1964zz}
\bibinfo{author}{\bibfnamefont{M.}~\bibnamefont{Bolsterli}} \bibnamefont{and}
  \bibinfo{author}{\bibfnamefont{E.}~\bibnamefont{Jezak}},
  \bibinfo{journal}{Phys. Rev.} \textbf{\bibinfo{volume}{135}},
  \bibinfo{pages}{B510} (\bibinfo{year}{1964}).

\bibitem[{\citenamefont{L{\"u}scher and Wolff}(1990)}]{Luscher:1990ck}
\bibinfo{author}{\bibfnamefont{M.}~\bibnamefont{L{\"u}scher}} \bibnamefont{and}
  \bibinfo{author}{\bibfnamefont{U.}~\bibnamefont{Wolff}},
  \bibinfo{journal}{Nucl. Phys.} \textbf{\bibinfo{volume}{B339}},
  \bibinfo{pages}{222} (\bibinfo{year}{1990}).

\end{thebibliography}

\clearpage
%
%
%
%
\begin{table}[!t]
\caption{
Simulation parameters for gauge configuration generation at 
($\kappa_{ud}$, $\kappa_s$)=(0.1376314,0.1367299).
The definition of parameters is same as in Ref.~\cite{Aoki:2008sm}.
\label{tab:conf_gene}
}
\begin{ruledtabular}
\begin{tabular}{c|cccc|cc}
$L^3 \times T$& \multicolumn{4}{c|}{$48^3\times 48$} & 
\multicolumn{2}{c}{$64^3\times 64$}\\\hline
\# run         & \multicolumn{4}{c|}{4} & \multicolumn{2}{c}{2} \\
$(N_0,N_1,N_2,N_3)$ & \multicolumn{4}{c|}{(2,2,2,6)} & 
\multicolumn{2}{c}{(2,2,2,8)} \\
Block size & \multicolumn{4}{c|}{$12^2\times 6^2$} & 
\multicolumn{2}{c}{$8^3\times 4$}\\
$\rho$       & \multicolumn{4}{c|}{0.998} & 
\multicolumn{2}{c}{0.998}\\
$N_{\rm poly}$ & \multicolumn{4}{c|}{320} & \multicolumn{2}{c}{340} \\
MD time       & 1000 & 1000 & 1000 & 1000 & 800 & 800 \\
$P_{acc}$(HMC)& 0.818 & 0.801 & 0.814 & 0.794 & 0.902 & 0.880 \\
$P_{acc}$(GMP)& 0.959 & 0.959 & 0.962 & 0.962 & 0.954 & 0.967 \\
\end{tabular}
\end{ruledtabular}
\end{table}
\begin{table}[!t]
\caption{
Number of configurations, 
separation of trajectories between each measurement in the units of $\tau$,
bin size in jackknife analysis in the units of configuration,
number of measurements on each configuration,
exponential smearing parameter set ($A,B$) in Eq.~(\ref{eq:smear}),
pion mass $m_\pi$ and nucleon mass $m_N$ are summarized
for each lattice size.
Number of measurements include factor four by
measurement with all four directions, which is explained in the text.
\label{tab:conf_meas}
}
\begin{ruledtabular}
\begin{tabular}{ccccccccc}
$L$ & $T$ & \# config. & $\tau_{\mathrm{sep}}$ 
& bin size & \# meas. & $(A,B)$
& $m_\pi$ [GeV] & $m_N$ [GeV]\\
\hline
48 & 48 & 400 & 10 & 20 & 1152 & (0.8,0.16) & 0.3001(14) & 1.057(2) \\
64 & 64 & 160 & 10 & 16 & 2048 & (0.8,0.16) & 0.2987(9)  & 1.053(2) \\
\end{tabular}
\end{ruledtabular}
\end{table}
\begin{table}[!t]
\caption{
\label{tab:dE_He}
Energy shift $-\Delta E_L$ in physical units 
and fit range for $^4$He and $^3$He channels 
on each spatial volume.
Extrapolated results in the infinite spatial volume limit 
are also presented. The first and second errors are
statistical and systematic, respectively.
}
\begin{ruledtabular}
\begin{tabular}{ccccc}
& \multicolumn{2}{c}{$^4$He}&
\multicolumn{2}{c}{$^3$He}\\
$L$  & $-\Delta E_L$ [MeV] & fit range & $-\Delta E_L$ [MeV] & fit range \\
\hline
48 & 46(13)($^{+55}_{-11}$) & 9--12 & 18.7(1.9)($^{+15}_{-3.1}$) &  8--12 \\
64 & 47(9)($^{+6}_{-8}$)    & 9--12 & 23.7(1.6)($^{+13}_{-2.5}$) &  9--12 \\
$\infty$ & 47(7)($^{+20}_{-11}$) & --- &
21.7(1.2)($^{+13}_{-1.6}$) & --- \\
\end{tabular}
\end{ruledtabular}
\end{table}
\begin{table}[!t]
\caption{
\label{tab:dE_NN}
Same as Table~\ref{tab:dE_He} for $^3$S$_1$ and $^1$S$_0$ channels.
}
\begin{ruledtabular}
\begin{tabular}{ccccc}
& \multicolumn{2}{c}{$^3$S$_1$}&
\multicolumn{2}{c}{$^1$S$_0$}\\
$L$  & $-\Delta E_L$ [MeV] & fit range & $-\Delta E_L$ [MeV] & fit range \\
\hline
48 & 13.8(0.9)($^{+3.6}_{-1.7}$) &  9--13 & 7.7(0.9)($^{+2.4}_{-1.2}$) &  9--13 \\
64 & 15.6(1.2)($^{+1.0}_{-1.3}$) & 10--14 & 9.5(0.9)($^{+0.5}_{-0.8}$) & 10--14 \\
$\infty$ & 14.5(0.7)($^{+2.4}_{-0.8}$) & --- &
8.5(0.7)($^{+1.6}_{-0.5}$) & --- \\
\end{tabular}
\end{ruledtabular}
\end{table}

\clearpage

%
%
%
%

\end{document}